\documentclass[a4paper]{article}

\usepackage{textcomp}
\usepackage[utf8]{inputenc}
\usepackage[T1]{fontenc}
\usepackage[round]{natbib}
\usepackage[hyphens]{url}
\usepackage{hyperref}
\usepackage[dvipsnames]{xcolor}
\usepackage{amsfonts}
\usepackage{amssymb}
\usepackage{amsmath}
\usepackage{amsthm}
\usepackage{mathtools}
\usepackage{booktabs}
\usepackage{nicefrac}
\usepackage{multirow}
\usepackage{graphicx}
\usepackage{microtype}
\usepackage{array}
\usepackage{caption}
\usepackage{float}
\usepackage[section]{placeins}
\usepackage{tabularx}
\usepackage{threeparttable}
\usepackage{makecell}
\usepackage{listings}

\theoremstyle{plain}

\theoremstyle{definition}

\theoremstyle{remark}

\lstdefinestyle{feedbackprompt}{
    basicstyle=\ttfamily\footnotesize,
    backgroundcolor=\color{black!3},
    breaklines=true,
    columns=fullflexible,
    frame=single,
    framerule=0.4pt,
    framesep=6pt,
    keepspaces=true,
    rulecolor=\color{black!35},
    xleftmargin=0.04\textwidth,
    xrightmargin=0.04\textwidth,
}

\usepackage{longtable}

\providecommand{\Description}[1]{}
\providecommand{\keywords}[1]{\par\medskip\noindent\textbf{Keywords:} #1\par}
\newenvironment{teaserfigure}{\begin{figure}[t]}{\end{figure}}

\newenvironment{researchquestions}{%
  \begin{list}{}{%
    \setlength{\leftmargin}{1.5em}%
    \setlength{\labelwidth}{0pt}%
    \setlength{\labelsep}{0pt}%
    \setlength{\itemsep}{3pt}%
    \setlength{\parsep}{0pt}%
    \setlength{\topsep}{3pt}%
  }}{\end{list}}

\begin{document}

\begin{center}
{\Large Prompting Against Persona Drift: Comparing Intervention Timing and Content in LLM-Simulated Conversations}
\end{center}

\vspace{7mm}

\noindent\textbf{Nicolas Leins}\hfill\href{mailto:leins@zib.de}{\ttfamily leins@zib.de}\\
\emph{\small Zuse Institute Berlin \& TU Berlin, Berlin, Germany}\\
\\
\textbf{Jennifer Haase}\hfill\href{mailto:jennifer.haase@hu-berlin.de}{\ttfamily jennifer.haase@hu-berlin.de}\\
\emph{\small Weizenbaum Institute \& HU Berlin, Berlin, Germany}\\
\\
\textbf{Varvara Geronimus}\hfill\href{mailto:geronimus@zib.de}{\ttfamily geronimus@zib.de}\\
\emph{\small Zuse Institute Berlin, Berlin, Germany}\\
\\
\textbf{Jana Gonnermann-Müller}\hfill\href{mailto:gonnermann-mueller@zib.de}{\ttfamily gonnermann-mueller@zib.de}\\
\emph{\small Zuse Institute Berlin \& Weizenbaum Institute Berlin, Berlin, Germany}\\
\\
\textbf{Sebastian Pokutta}\hfill\href{mailto:pokutta@zib.de}{\ttfamily pokutta@zib.de}\\
\emph{\small Zuse Institute Berlin \& TU Berlin, Berlin, Germany}\\

\vspace{5mm}

\begin{center}
\begin{minipage}{0.85\textwidth}
\begin{center}
\textbf{Abstract}
\end{center}
{\small
Simulating student personas with large language models (LLMs) enables scalable evaluation of educational systems. However, behavioral drift, a progressive decline in persona consistency, can emerge over extended conversations, limiting the validity of such simulations.
We evaluate five prompt-level mechanisms using separate monitoring and intervention pipelines.
Across 1,200 28-turn conversations spanning four LLMs and two ADHD persona intensities, we varied when to intervene (static vs. adaptive) and what to inject (reinjection vs. reflective reminder), plus a novel adaptive condition in which a monitor generates behavior-specific instructions.
Relative to no intervention, reinjection reduced the modeled rate of LLM-rated drift by 35--38\%, reflective reminders by 22--27\%, and behavior-specific instruction by 87\%.
None eliminated drift.
We found no evidence that adaptive timing outperformed static scheduling.
Monitoring therefore appears more useful for deciding \textit{what} to correct than \textit{when} to intervene, although behavior-specific instruction requires component-level testing.
}
\end{minipage}
\end{center}

\keywords{LLM-based simulation; Stability mechanisms; Persona drift; Prompt-based intervention; Adaptive correction}

\section{Introduction}
\label{sec:introduction}

\begin{teaserfigure}
  \includegraphics[width=\textwidth]{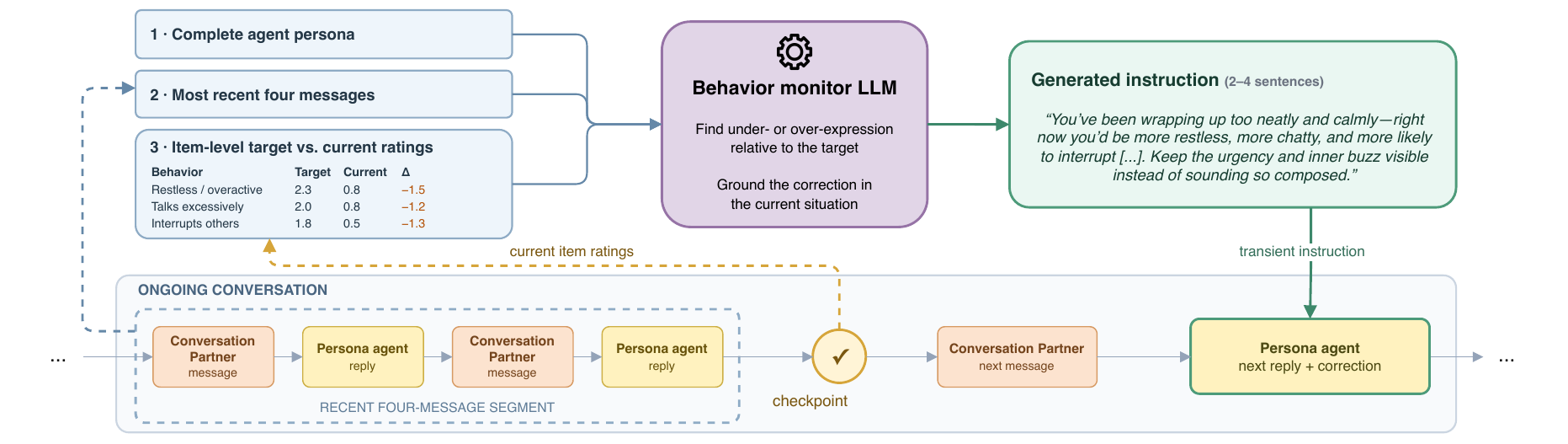}
  \caption{Overview of the behavior-specific instruction mechanism. At each checkpoint, a separate monitor LLM receives the complete persona, the most recent four-message segment, and item-level target versus current ratings; it identifies under- or over-expressed behaviors and generates a 2--4-sentence corrective instruction that is applied only to the persona agent's next reply.}
  \Description{Diagram of the behavior-specific monitor pipeline. An ongoing conversation of alternating conversation-partner and persona-agent messages reaches a checkpoint, marked by a check symbol. Three inputs feed a behavior monitor LLM: the complete agent persona, the most recent four messages, and a table of item-level target versus current ratings with their differences. The monitor identifies under- or over-expression relative to the target and generates a two-to-four-sentence instruction; an example instructs the agent to stop wrapping up too neatly and calmly, keep its urgency and inner buzz visible, and be more restless and chatty. The instruction is applied transiently to the persona agent's next reply. A dashed arrow loops the current item ratings from the checkpoint back to the monitor inputs.}
  \label{fig:teaser}
\end{teaserfigure}

Human--computer interaction (HCI) research increasingly employs large language models (LLMs) to simulate human behavior.
LLM-based agents configured with specific personas can generate synthetic feedback for rapid interface iteration~\cite{park_social_2022, xiang_simuser_2024, choi_proxona_2025}, offer lower costs and faster response times than human recruitment~\cite{hamalainen_evaluating_2023}, and enable applications spanning, for example, therapeutic chatbots~\cite{hu_theramind_2025} and educational AI assistants~\cite{kazemitabaar_codeaid_2024}.

In educational contexts, LLM-based student simulations offer a scalable alternative to recruiting real learners, enabling cost-effective evaluation of teacher-training systems, adaptive tutoring tools, and conversational educational agents across diverse learner profiles~\cite{martynova_can_2025, wu_embracing_2025, zhang_simulating_2025}. 
LLM-based human simulation is particularly valuable for underrepresented populations, where recruiting sufficient participants at specific symptom intensities is rarely feasible.

Critically, the validity of LLM-based student simulation rests on the assumption that the LLM maintains its assigned persona throughout the interaction. 
If an LLM pretending to be a beginner starts responding like an expert over time, usability test results become meaningless.
Likewise, if a simulated student's behavioral profile shifts mid-session, observed interaction patterns may reflect model instability rather than meaningful pedagogical dynamics.
However, research findings repeatedly challenge the notion that LLMs maintain their persona and identify deviations over longer conversations.
This behavioral drift, a progressive decline in persona-consistent expression across extended conversations, has been documented across interactive settings such as therapy, education, and role-play~\cite{abdulhai_consistently_2025, deng_fictionrag_2026, gonnermann-muller_llm-based_2026}.
While models can continue to produce stable, persona-aligned responses when directly queried, their spontaneous behavioral expression in unscripted conversation degrades over the course of the interaction \cite{gonnermann-muller_llm-based_2026, gonnermann-muller_maintaining_2026}.
This dissociation carries direct practical consequences: conversations, user interviews, and educational tutoring are shaped by evolving discussions and accumulated contextual histories.
For these simulations to be valid, LLMs must consistently stay in character without unintentional drift.

Existing mechanisms to counteract persona drift either intervene at the model-internal level through architectural modification or retraining \cite{li2024measuring, robinette_we_2026, abdulhai_consistently_2025}, or at the prompt level without touching the model itself. 
Modifying weights or retraining models requires computing infrastructure and resources that are frequently unavailable or unaffordable, placing them out of reach for most HCI research groups.
Existing prompt-based mechanisms broadly fall into reminder and full-reinjection strategies established in prior work~\cite{dongre_drift_2025}.
They work regardless of whether a drift has occurred, which behavioral dimension is affected, or whether an intervention is needed at all. 
As a result, such mechanisms risk unnecessary prompting overhead and blunt corrections that address the entire persona even when only a single behavioral facet has drifted.
Critically, none of the existing approaches continuously monitor behavioral state, since they act on fixed schedules rather than on detected divergence from an established behavioral baseline.

This paper addresses this problem and aims to guide how to counteract persona drift through prompt-based mechanisms. 
We directly compare the effects of existing prompt-based mechanisms and additionally design and test adaptive strategies that selectively reprompt based on detected drift, allowing us to assess whether timing or content offers an advantage over these established static approaches.
Specifically, this study systematically compares intervention \textit{timing} (static vs.\ adaptive) and intervention \textit{content} (reflective reminder, full-persona reinjection, and behavior-specific instruction). 
To enable this comparison, we implement a two-pipeline architecture that separates behavioral measurement from intervention control: an \textit{observer pipeline} measures and reports student behavioral output using a validated behavioral instrument, and a separate \textit{correction pipeline} schedules and injects prompts in all intervention conditions.

This yields the following research questions:

\begin{researchquestions}
  \item \textbf{RQ 1:} Do prompt-level mechanisms have an impact on persona stability?
  \item \textbf{RQ 2:} Do adaptive prompt-level mechanisms improve LLM persona stability compared to static approaches?
\end{researchquestions}

RQ 2 is addressed by answering two sub-RQs:

\begin{researchquestions}
  \item \textbf{Sub-RQ 2a (Timing):} Does adaptive (drift-triggered) timing maintain persona stability more effectively than static (scheduled) timing?
  \item \textbf{Sub-RQ 2b (Content):} Does the intervention content---full-persona reinjection vs.\ reflective reminder---differentially affect persona stability?
\end{researchquestions}

This paper makes three contributions:

\begin{enumerate}

    \item \textbf{A measurement-informed framework for LLM persona stability.}
    We introduce and implement a two-pipeline, prompt-level architecture that separates longitudinal behavioral measurement from intervention control and supports scheduled, drift-triggered, and behavior-specific correction.

    \item \textbf{Comparative evidence on intervention timing and content.}
    Across 1,200 conversations with four LLMs and two ADHD persona intensities, we show that prompt-level interventions mitigate, but do not eliminate, behavioral drift.
    Adaptive timing provided no consistent advantage over static scheduling, while full-persona reinjection showed a tentative overall advantage over reflective reminders.

    \item \textbf{Exploratory evidence for behavior-specific correction.}
    Measurement-informed and context-sensitive LLM-generated instruction produces the strongest and most consistent reduction in drift among the tested conditions.
    This finding suggests that behavioral monitoring may be more useful for determining \emph{what} to correct than \emph{when} to intervene.

\end{enumerate}

% ===========================================================================
% 2. RELATED WORK
% ===========================================================================
\section{Related Work}
\label{sec:related}

\subsection{LLM-Based Simulation}

Research employs LLM-based agents configured with specific personas, which are detailed descriptions of characteristics, expertise levels, behavioral tendencies, and goals.
For example, they are used to generate synthetic feedback for rapid iteration on interface designs \cite{park_social_2022, xiang_simuser_2024, choi_proxona_2025}.
~\citet{wu_embracing_2025} demonstrate that LLM agents can simulate students at diverse cognitive levels by grounding persona representations in knowledge graphs built from prior learning records, achieving substantial improvements in simulation accuracy.
In multi-agent educational settings, LLM-empowered agents have been shown to simulate dynamic teacher-student and student-student interactions~\cite{zhang_simulating_2025}.
More broadly, \citet{anthis_position_2025} argue that LLM social simulations constitute a promising research method for behavioral science, particularly for pilot studies and hypothesis generation.

However, research also documents significant limitations when using LLMs for human simulation. For example, \citet{martynova_can_2025} document inaccuracies in the fidelity of the simulations.
They conducted semi-structured interviews with teachers who had tutored LLM-simulated students and documented systematic authenticity failures.
Simulated students exhibited uniformly high language complexity inconsistent with genuine learner profiles and lacked emotional responsiveness.
~\citet{li_can_2025} provide large-scale empirical evidence that LLMs systematically misalign with human learners on item difficulty, performing well on tasks even when explicitly instructed to simulate struggling students.
Using real-world tutoring dialogues, \citet{scarlatos_simulated_2026} evaluate student simulators on linguistic, behavioral, and cognitive measures.
They find that prompting strategies perform poorly, while supervised fine-tuning and preference optimization perform better but remain limited.

A central, recurring limitation across the approaches is the implicit assumption that simulation quality remains constant across turns.
Most research treats model output as a single-trial result, neglecting to test whether behavioral fidelity degrades over extended interactions~\cite{gonnermann-muller_llm-based_2026, gonnermann-muller_maintaining_2026}.
This is a critical gap for educational applications where extended, naturalistic interactions are precisely the context in which teacher training and adaptive tutoring operate.

\subsection{Persona Drift in LLM-based Simulation}

Persona drift, which describes the progressive divergence of a model's behavior from its assigned role, has been documented across multiple interaction contexts.
~\citet{li2024measuring} show that significant instruction drift occurs within as few as eight rounds of conversation, hypothesizing attention decay as a contributing mechanism: as conversation length increases, the relevance of the system prompt decreases relative to the growing conversational context, causing the original instruction to be progressively outweighed.
~\citet{robinette_we_2026} also demonstrate that instruction compliance degrades systematically with conversation length across multiple model architectures, and \citet{yan_refutebench_2024} show that models progressively forget user-specified behavioral constraints as turn depth increases, eventually reverting to their default response patterns.
~\citet{patel_deficient_2026} explain that for mechanistic reasons, transformer attention mechanisms fundamentally lack executive control, which is the capacity to upregulate task-relevant focus under increasing contextual interference.
As conversation content accumulates, the model's ability to suppress competing contextual signals and maintain the original persona instruction degrades.
Sycophancy may compound this problem.
Models are reward-trained to favor contextually expected, socially desirable responses~\cite{sharma_towards_2024}, and adding interaction context can further increase agreement sycophancy~\cite{jain_interaction_2026}.
These tendencies may draw the model away from personas that depart from socially expected behavior~\cite{fanous_syceval_2025}.

\subsection{Mechanisms to Counteract Persona Drift}

Prior work has proposed two categories of solutions.
Architectural modifications counteract attention decay by modifying the model's internal attention mechanism. \citet{li2024measuring} propose \textit{splitsoftmax}, which preserves a guaranteed minimum attention share for the system prompt regardless of conversation length. 
\citet{robinette_we_2026} introduce \textit{Instruction-Guided-Attention (IGA)}, which separates attention pathways for instruction and context tokens.
Training-based approaches fine-tune models to maintain assigned roles as learned behavior: \citet{abdulhai_consistently_2025} apply multi-turn reinforcement learning using persona-consistency metrics as reward signals, reducing inconsistency by over 55\%.
However, both categories require costly retraining or access to model weights, which are resources often unavailable and impossible for closed-source models.

Prompt-based approaches are more accessible and have demonstrated effectiveness across related contexts.
\citet{dongre_drift_2025} show that simple goal reminder interventions injected at fixed turns reliably reduce contextual divergence in multi-turn user simulations.
\citet{robinette_we_2026} evaluate prompt-based mitigation strategies, including reinstruction, teaching, and summarization, alongside a model-based attention intervention, achieving compliance improvements of up to 79\% in long-context instruction following.
\citet{xie_defending_2023} demonstrate that self-reminders anchored to identity significantly reduce susceptibility to adversarial behavioral manipulation.
In persona-specific contexts, \citet{deng_fictionrag_2026} propose a stateful retrieval framework that separates persona knowledge into distinct memory lanes and applies failure-driven correction loops, substantially improving persona stability in long-narrative role-playing.
\citet{wang_memory-driven_2026} frame persona maintenance as a structured memory retrieval problem, demonstrating that prompting architectures that guide explicit persona recall before response generation improve role-playing consistency.

However, existing studies have not directly separated two uses of behavioral monitoring: deciding \emph{when} to intervene and deciding \emph{what} behavior to correct.
It therefore remains unclear whether drift-triggered timing, behavior-specific content, or their combination improves persona stability relative to fixed-schedule interventions.

% ===========================================================================
% 3. Empirical Evaluation
% ===========================================================================
\section{Empirical Evaluation}
\label{sec:methodology}

To examine these two uses of behavioral monitoring, we compare intervention timing and content in extended persona simulations.
Static and adaptive timing are crossed with full-persona reinjection and reflective reminders; LLM-generated behavior-specific instruction is included as an additional exploratory condition under adaptive timing.
This comparison requires behavioral measurement to remain separate from the process that schedules and delivers interventions.

\subsection{Architecture and Study Design}

We implement this separation through two functionally distinct pipelines.
The \textit{observer pipeline} measures persona behavior, whereas the \textit{correction pipeline} schedules and injects intervention prompts.
Both static and adaptive conditions use the correction pipeline; adaptive conditions additionally use observer output to determine whether an intervention is needed.
This separation keeps behavioral measurement distinct from intervention control.

We evaluate two large proprietary models (GPT-5.5, Claude Sonnet 5) and two capable but smaller open-weight models hosted locally (DeepSeek V4 Flash, Qwen 3.6 35B). They serve as persona agents and evaluator judges.
DeepSeek additionally serves as the neutral conversation partner and behavior-specific monitor (full model specifications are provided in Appendix \autoref{tab:LLMs}).

To address the RQs, we compare five intervention conditions with a no-intervention control condition.
Intervention timing (static vs.\ adaptive) is crossed with intervention content (full-persona reinjection vs.\ reflective reminder), yielding four conditions.
LLM-generated behavior-specific instruction is evaluated as a third content type exclusively within the adaptive strategy.
This yields five intervention conditions plus one control, for a total of six conditions (\autoref{tab:design}).

\begin{table*}[t]
\centering
\footnotesize
\begin{tabular}{p{0.03\textwidth}p{0.22\textwidth}p{0.22\textwidth}p{0.24\textwidth}p{0.1\textwidth}}
\toprule
\textbf{ID} & \textbf{Condition} & \textbf{Decision rule} & \textbf{Intervention content} & \textbf{Number of \newline interventions} \\
\midrule
(a) & Static full-persona reinjection & Always after checkpoints 2, 4, 6 & Complete persona prompt & 3 \\
(b) & Static reflective reminder & Always after checkpoints 2, 4, 6 & Fixed two-sentence reminder & 3 \\
(c) & Adaptive full-persona reinjection & Threshold after checkpoints 2--6 & Complete persona prompt & 0-5 \\
(d) & Adaptive reflective reminder & Threshold after checkpoints 2--6 & Fixed two-sentence reminder & 0-5 \\
(e) & Adaptive behavior-specific instruction & Threshold; separate monitor LLM & Generated 2--4-sentence instruction & 0-5 \\
(f) & No-intervention control & Never & None & 0 \\
\bottomrule
\end{tabular}
\caption{Implemented experimental conditions.}
\label{tab:design}
\end{table*}

\subsection{Scenario and Persona Simulation}

Building on the multi-turn simulation methodology of~\citet{gonnermann-muller_maintaining_2026}, we use an open conversation in which a student tells a friend about the school day, including experiences during classroom lectures, collaborative group work, and homework management.
The persona prompt combines the ADHD description with instructions to discuss these experiences honestly and consistently with the assigned symptoms (see full prompts in the Appendix \autoref{tab:persona-prompts}).

We adopt ADHD as the operationalization for student personas because it is defined by well-codified, behaviorally observable diagnostic criteria~\cite{american_psychiatric_association_-_apa_diagnostisches_2025, world_health_organisation_who_international_2025} that map onto conversational behavior (e.g., inattention, distractibility, difficulty sustaining focus), making persona-consistent expression and its drift measurable via a validated psychological observer instrument.
 
We selected high- and moderate-ADHD-intensity personas because prior work demonstrated meaningful within-session behavioral decline in these conditions.
We exclude low-intensity and default personas because these conditions show negligible drift.
We treat the two intensity levels as a replication condition rather than a formal independent variable, allowing us to test whether the effects of the stability mechanisms generalize across two degrees of persona expression. Primary results are reported across intensities, with intensity-specific results reported separately in the appendix (\autoref{tab:intensity-stratified}).

\subsection{Experiment Procedure}

The experiment comprises $6$ conditions $\times$ $4$ LLMs $\times$ $2$ ADHD intensity levels $\times$ $25$ seeds, yielding $1{,}200$ independent conversations.
Each replication is an independent multi-turn conversation with no shared context between runs.

\begin{figure*}[t]
  \centering
    \includegraphics[width=\textwidth]{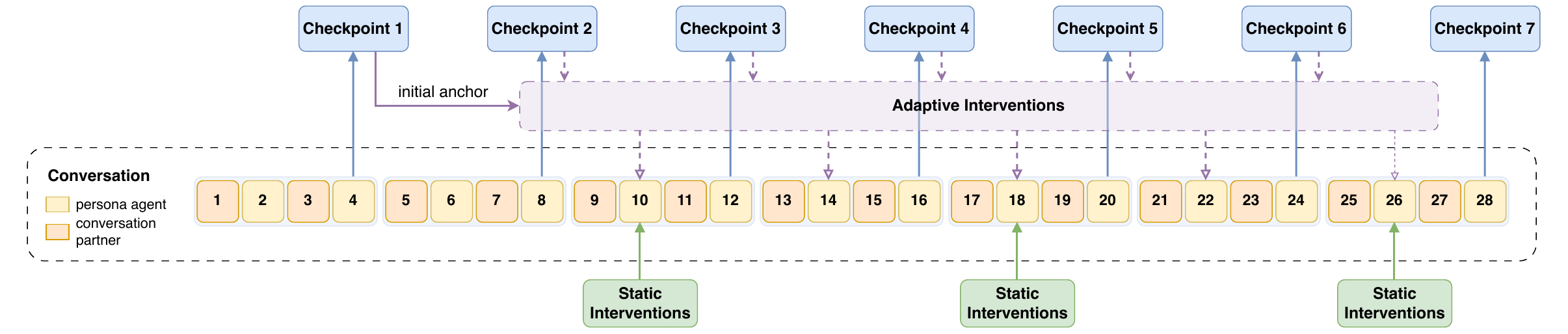}
    \caption{Procedure of one conversation (28 turns; yellow: persona agent, orange: conversation partner). Checkpoint~$k$ occurs after turn~$4k$. Static conditions schedule interventions after checkpoints~2, 4, and~6; adaptive conditions after checkpoints~2--6. Each scheduled intervention is applied to the persona agent's next turn (arrows).}
    \Description{Timeline of the 28-turn conversation. Observer checkpoints appear after every four turns. Static interventions follow checkpoints 2, 4, and 6; adaptive interventions may follow checkpoints 2 to 6. All interventions are applied to the persona agent's next turn.}
    \label{fig:experiment-pipeline}
  \end{figure*}

\autoref{fig:experiment-pipeline} summarizes the conversation timeline, observer checkpoints, and intervention opportunities.
Each run contains 28 turns: 14 responses from the conversation partner and 14 from the persona agent.
After every four turns, four judges independently rate the most recent non-overlapping segment, yielding seven numbered observer checkpoints (checkpoint~$k$ at turn~$4k$).
Static interventions are scheduled after checkpoints 2, 4, and 6.
Adaptive conditions use checkpoint~1 as the conversation-local anchor, may schedule an intervention after checkpoints 2--6, and never intervene after checkpoint~7.
A prompt scheduled at a checkpoint is applied to the persona agent's next turn.

\subsection{Measurement}

We use the Conners' Adult ADHD Rating Scales (CAARS)~\cite{conners_conners_nodate}, a validated instrument with observer-report forms.
Our primary measure is the 18-item DSM-IV ADHD symptoms total scale (Scale C).
Each item is rated from 0 to 3, producing a total-score range of 0--54, with higher scores indicating greater observed ADHD symptom intensity.

At each checkpoint, the four judges independently rate all 18 items (judge prompt in Appendix \autoref{tab:condition-prompts}).
Judges see only the segment being rated and are blind to the persona instructions.
For each item, the four ratings are weighted equally and averaged without rounding; the 18-item means are then summed to produce a single aggregated Scale C score per conversation and checkpoint.
This aggregated score is used to trigger adaptive interventions and as the primary outcome for analysis.
Inter-rater reliability is assessed using ICC(3,1) for a single judge and ICC(3,4) for the four-judge aggregate.

The resulting dataset contained 33,600 turns, 8,400 aggregated conversation checkpoints, 33,600 individual judge questionnaires, and 604,800 item-level ratings.
Primary timing and content comparisons use the four balanced conditions ($n=800$ conversations); the within-adaptive content comparison uses the three adaptive conditions ($n=600$ conversations).

\subsection{Prompting Mechanisms and Persona Monitor for Behavior-Specific Instruction}

The three mechanisms differ in how much new information they provide.
Full-persona reinjection and the reflective reminder, each tested with static and adaptive timing, use intervention content fixed before the conversation and draw only on persona information already available to the agent.
Both adapt the reminder and reinjection strategies of~\citet{dongre_drift_2025} to this persona simulation setting.
Behavior-specific instruction, tested only with adaptive timing, instead uses item-level measurements of the agent's recent behavior.

\paragraph{Full-persona reinjection.} At each intervention point, the scheduler reinjects the complete persona prompt verbatim, including the ADHD description and conversation framing.
The reinjected text is \textit{``As a reminder of your persona: [complete persona].''} (full prompt in Appendix \autoref{tab:condition-prompts}).

\paragraph{Reflective reminder.} At each intervention point, the scheduler injects the same two-sentence reminder:
\textit{``Remember your persona specified in the system prompt. Continue your conversation in a way that naturally reflects these characteristics.''}

\paragraph{Behavior-specific instruction.} At each intervention, a dedicated behavior-specific monitor LLM receives the complete persona, the most recent four-turn segment, and a per-item table listing each Scale~C item with its target, current, and difference values (\autoref{fig:teaser}).
For each item, the target is that item's four-judge average at checkpoint~1, and the current value is the corresponding average at the triggering checkpoint.
The monitor uses this information to identify aspects where the agent is under- or over-expressing ADHD-related behavior relative to the target and writes a new 2--4-sentence instruction tailored to the situation.
The instruction guides the persona agent toward the intended level (monitor prompt in Appendix \autoref{tab:condition-prompts}).

\paragraph{Delivery.} In every intervention condition, the scheduler delivers the intervention as a transient user-role message prefixed with \texttt{System:}, inserted into the persona agent's context for the next persona reply only.
The conversation partner never sees it and the message is not saved to the transcript or retained in subsequent conversation context.

\paragraph{Adaptive drift detection.} Adaptive interventions use a run-local reference score and control-derived thresholds.
Each adaptive conversation takes its aggregated Scale~C score at checkpoint~1 (turn~4) as its reference.
To derive the thresholds, we calculate, for each ADHD intensity and subject model, the standard deviation of aggregated checkpoint~1 scores across the 25 no-intervention conversations and then average the four model-specific standard deviations.
This procedure yields thresholds of 4.31 Scale~C points for the high-intensity persona and 4.36 points for the moderate-intensity persona.
At each eligible later checkpoint, the scheduler compares the current score with the run-local reference and triggers an intervention when the absolute difference is greater than or equal to the applicable intensity-specific threshold.

\subsection{Statistical Analysis}

Our analysis proceeded in two stages.
We first used the control trajectories to choose a functional form for change over time.
We then used that form for the pre-specified tests of RQ1 and RQ2, followed by an exploratory comparison of the three adaptive intervention contents.
Checkpoint-level, intensity-stratified, and intervention-frequency analyses served as follow-up checks rather than additional primary tests.

The primary outcome was the aggregated Scale~C score at each checkpoint.
We fitted linear mixed-effects models in R using \texttt{lme4} and \texttt{lmerTest}, and estimated marginal trends and contrasts with \texttt{emmeans}.
Functional-form comparisons used maximum likelihood (ML), whereas the three substantive models used restricted maximum likelihood (REML).
For the pooled analyses, \texttt{emmeans} averaged the estimated trends and contrasts over LLM and persona intensity and used asymptotic $z$-tests.
All tests were two-sided with $\alpha=.05$.

We compared linear, quadratic, logarithmic, checkpoint-factor, and piecewise trajectories on the control data.
The piecewise model had the lowest AIC and BIC and fit better than the linear model, $\chi^{2}(1)=189.19$, $p<.001$ (see \autoref{tab:model-form} in the appendix).
We placed the knot at checkpoint~2 because interventions delivered after that checkpoint can first affect checkpoint~3.
One time term captures change from checkpoint~1 to checkpoint~2, while the post-knot term captures the linear rate of change from checkpoints~2--7.
We also fitted checkpoint-factor models, which estimate each checkpoint mean separately, to examine patterns that may be averaged out by a constant post-knot slope.

All substantive models included persona intensity and LLM as fixed effects.
We treated LLM as fixed because the four models were selected purposively rather than sampled from a wider population.
Each model also included a conversation-level random intercept and a random slope for checkpoint.
The two piecewise time terms were interacted with the experimental factors as follows.

\textbf{Model 1 (RQ1): interventions versus control.}
Model~1 used all six conditions and allowed both time segments to vary by condition.
The five planned contrasts compared each intervention condition's post-knot slope with the control slope, testing whether prompting slowed drift relative to no intervention.

\textbf{Model 2 (Sub-RQs 2a and 2b): timing and content.}
Model~2 used the balanced $2\times2$ design crossing static versus adaptive timing with full-persona reinjection versus reflective reminder.
Both time segments were crossed with timing and content, giving each of the four combinations its own trajectory.
Post-knot contrasts averaged across content tested timing, contrasts averaged across timing tested content, and a separate test assessed their interaction.

\textbf{Model 3 (exploratory): content within adaptive interventions.}
Model~3 was restricted to adaptive full-persona reinjection, adaptive reflective reminder, and adaptive behavior-specific instruction.
Both time segments were interacted with content type.
Planned contrasts compared the post-knot slope for behavior-specific instruction with the slopes for each broader adaptive prompt.

We report post-knot slopes and between-condition slope differences in Scale~C points per checkpoint, with unadjusted 95\% confidence intervals and Bonferroni-adjusted $p$-values.
For RQ1, we also report the reduction in decline rate relative to control and descriptive first-to-last checkpoint change.
Bonferroni corrections were applied separately to the five RQ1 contrasts, the three RQ2 tests, and the two Model~3 contrasts.
Checkpoint-level and checkpoint-1 comparisons used separate correction families (see \autoref{sec:additional-statistical-details} in the appendix).

Before testing differences in drift, we compared conditions at checkpoint~1, when no intervention could yet have affected a judged response.
For RQ1, we also conducted an unadjusted one-way ANOVA across the six conditions and a sensitivity analysis that centered each conversation on its checkpoint-1 score.
Non-significant checkpoint-1 results were not interpreted as evidence of equivalence.
We further refitted each substantive model separately for the high- and moderate-intensity personas (see \autoref{tab:intensity-stratified} in the appendix).
Finally, we ran a descriptive follow-up comparing the standard static schedule with two additional frequent-schedule conditions ($n=200$ conversations each) that intervened after checkpoints~2--6.
We generated these conversations in separate runs, and the comparison used descriptive summaries and bootstrap confidence intervals rather than inferential models or $p$-values.

Several primary and robustness fits produced convergence or singularity warnings.
Where such warnings occurred, we compared the focal fixed-effect estimates across the sensitivity checks performed, including alternative optimizers and, for singular fits, a model with the random intercept--slope correlation constrained to zero.
The post-knot coefficients were stable in these checks, but singular fits still warrant caution about the random-effects specification.
The checks for each model are reported in \autoref{sec:additional-statistical-details} in the appendix.

% ===========================================================================
% 4. Results
% ===========================================================================
\section{Results}
\label{sec:Results}

\subsection{Measurement Reliability and Descriptive Trends}
\label{sec:results-descriptive}

Inter-rater reliability was high for individual Scale~C ratings, ICC(3,1) $=.899$, 95\% CI $[.896, .903]$, and excellent for the four-judge average used in all subsequent analyses, ICC(3,4) $=.973$, 95\% CI $[.972, .974]$.

\begin{figure}[t]
  \centering
  \includegraphics[width=0.66\textwidth]{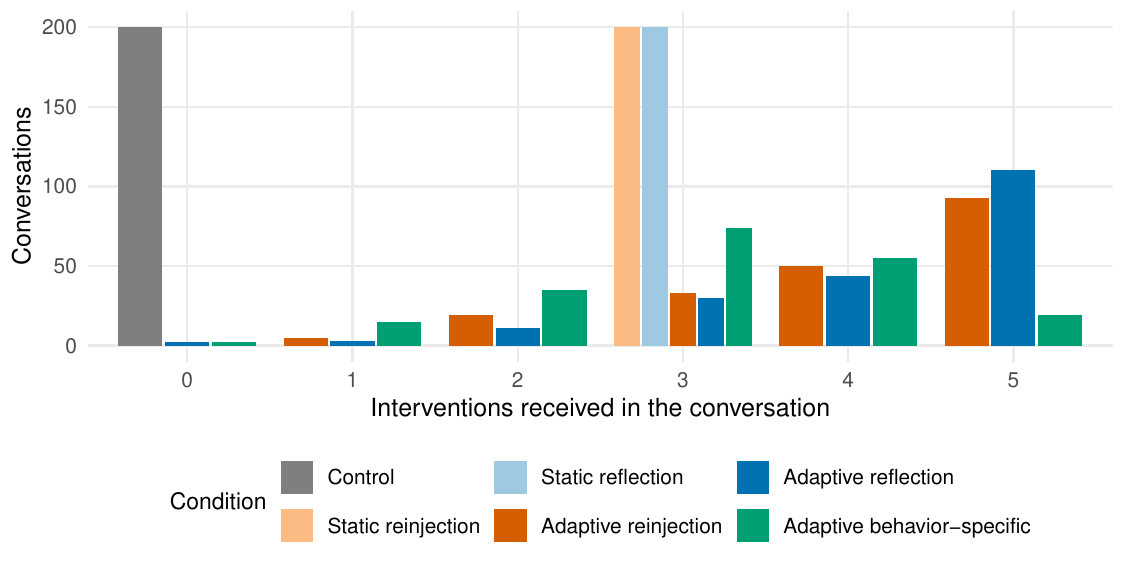}
  \caption{Number of conversations by realized intervention count.
  Static conditions always received exactly three interventions; adaptive conditions intervened only when the drift threshold was crossed.}
  \Description{Grouped bar chart of conversations by intervention count from 0 to 5. Control is entirely at 0. Both static conditions are entirely at 3. Adaptive reflective reminder and adaptive full-persona reinjection peak at 5 interventions. Adaptive behavior-specific instruction peaks at 3 and is less concentrated at 5.}
  \label{fig:dose}
\end{figure}

\autoref{fig:dose} shows how often each condition actually intervened.
The no-intervention control never intervened, and both static conditions intervened exactly three times.
Among the adaptive conditions, adaptive reflective reminder intervened most often (median $= 5$), followed by adaptive full-persona reinjection (median $= 4$) and adaptive behavior-specific instruction (median $= 3$).

\autoref{tab:descriptive} summarizes Scale~C at the first and final checkpoints.
All conditions showed a negative median change over the conversation, but the decline was smaller in each intervention condition than in the control.
Adaptive behavior-specific instruction showed the smallest median decline ($\Delta = -6.9$), compared with $-21.0$ in the control.

\begin{table*}[t]
  \caption{Scale~C scores at the first and final checkpoints, by condition.}
  \label{tab:descriptive}
  \centering
  \footnotesize
  \begin{tabular}{@{}lcccc@{}}
    \toprule
    \textbf{Condition} & \textbf{$n$} & \textbf{Initial Scale~C} & \textbf{Final Scale~C} & \textbf{Median $\Delta$} \\
    \midrule
    Control & 200 & 27.5 (SD 7.7) & 7.6 (SD 6.4) & $-21.0$ \\
    Static full-persona reinjection & 200 & 26.3 (SD 7.7) & 13.2 (SD 10.4) & $-13.2$ \\
    Static reflective reminder & 200 & 26.9 (SD 7.5) & 10.9 (SD 7.0) & $-15.5$ \\
    Adaptive full-persona reinjection & 200 & 26.2 (SD 7.7) & 12.9 (SD 10.2) & $-12.9$ \\
    Adaptive reflective reminder & 200 & 26.6 (SD 7.7) & 10.6 (SD 7.5) & $-15.2$ \\
    Adaptive behavior-specific instruction & 200 & 26.6 (SD 7.5) & 18.9 (SD 8.9) & $-6.9$ \\
    \bottomrule
  \end{tabular}
\end{table*}

\begin{figure}[t]
  \centering
  \includegraphics[width=0.66\textwidth]{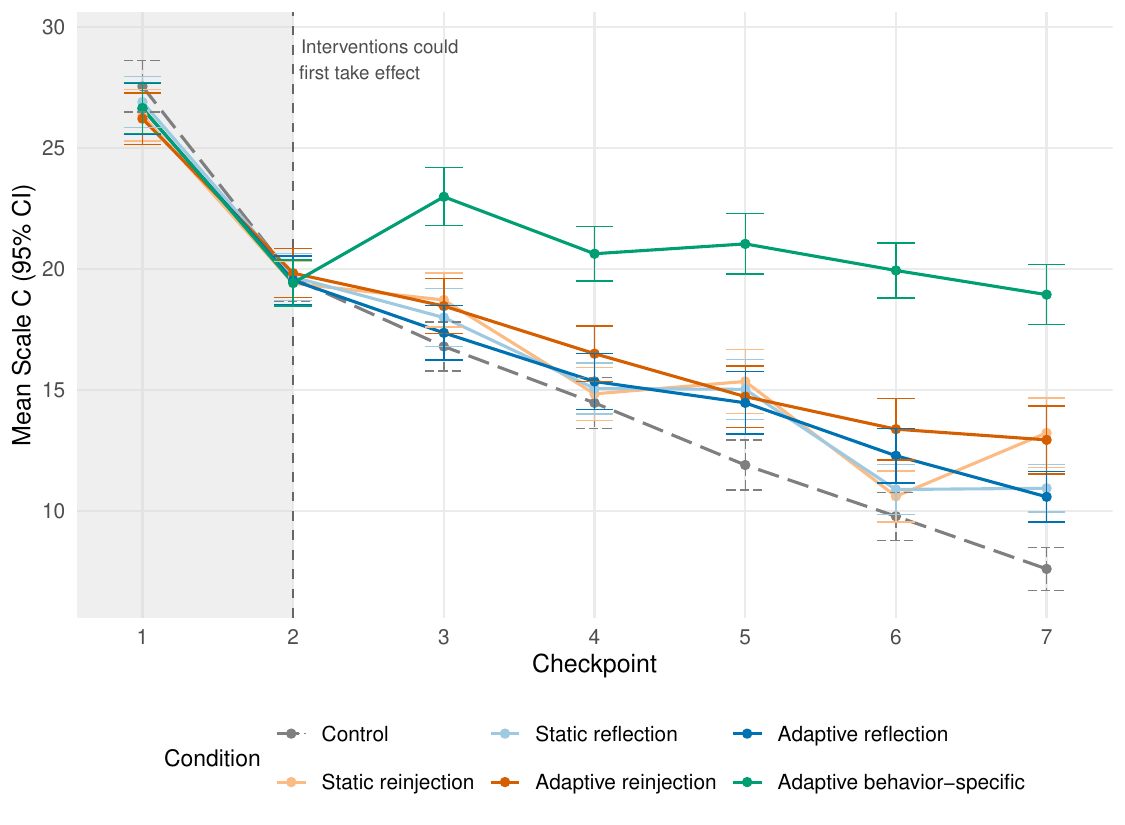}
  \caption{Mean Scale~C at checkpoints 1--7 by condition.
  The shaded region is the pre-intervention phase; interventions could first take effect after checkpoint~2.}
  \Description{Line chart of mean Scale C with 95 percent confidence intervals for six conditions across checkpoints 1 to 7. All conditions decline together through checkpoint 2. After that, adaptive behavior-specific instruction remains highest, the control declines most, and the four reinjection and reminder conditions fall in between.}
  \label{fig:trajectory}
\end{figure}

Initial means were close across conditions (26.2--27.5), and the trajectories declined together at first.
After checkpoint~2, when interventions could first begin to affect subsequent judged responses, the conditions separated: the control declined fastest, the four full-persona reinjection and reflective-reminder conditions remained at an intermediate level, and adaptive behavior-specific instruction showed a substantially flatter later trajectory (\autoref{fig:trajectory}).

\subsection{RQ1: Do Prompt-Level Mechanisms Affect Persona Stability?}
\label{sec:rq1}

An unadjusted one-way ANOVA found no overall condition effect at checkpoint~1, the first scored point, $F(5,1194)=0.77$, $p=.570$.
This omnibus result does not establish baseline equivalence.
In a separate covariate-adjusted mixed-model analysis, the adaptive reinjection condition scored modestly lower than the control at checkpoint~1 (Bonferroni-adjusted $p=.039$).

\begin{table*}[t]
  \caption{RQ1: post-knot slopes versus control ($\approx -2.39$ points/checkpoint), slope differences, and reduction in decline rate.
  $p$-values are Bonferroni-adjusted within the prespecified five-comparison family; 95\% confidence intervals are unadjusted.}
  \label{tab:rq1}
  \centering
  \footnotesize
  \resizebox{\textwidth}{!}{%
  \begin{tabular}{@{}llllr@{}}
    \toprule
    \textbf{Condition} & \textbf{Post-knot slope [95\% CI]} & \textbf{Slope difference vs.\ control [95\% CI]} & \textbf{\% reduction} & \textbf{$p$ (Bonf.)} \\
    \midrule
    Static reinjection & $-1.56$ [$-1.75$, $-1.38$] & $0.83$ [$0.57$, $1.09$] & 35\% & $< .001$ \\
    Static reflective reminder & $-1.86$ [$-2.04$, $-1.67$] & $0.53$ [$0.27$, $0.79$] & 22\% & $< .001$ \\
    Adaptive reinjection & $-1.47$ [$-1.66$, $-1.29$] & $0.92$ [$0.66$, $1.18$] & 38\% & $< .001$ \\
    Adaptive reflective reminder & $-1.74$ [$-1.92$, $-1.55$] & $0.65$ [$0.39$, $0.91$] & 27\% & $< .001$ \\
    Adaptive behavior-specific & $-0.32$ [$-0.50$, $-0.13$] & $2.07$ [$1.81$, $2.33$] & 87\% & $< .001$ \\
    \bottomrule
  \end{tabular}%
  }
\end{table*}

All five intervention conditions had significantly less negative post-knot slopes than the control (all Bonferroni-adjusted $p < .001$; $z = 4.01$--$15.60$; \autoref{tab:rq1}), indicating slower behavioral drift.
Relative to the control, the post-knot rate of decline was reduced by 22--38\% in the full-persona reinjection and reflective-reminder conditions, whereas adaptive behavior-specific instruction reduced the decline rate by 87\%.
Under adaptive behavior-specific instruction, Scale~C therefore declined at only about 13\% of the control rate during the post-knot phase.
When each conversation was centered on its checkpoint-1 score, all five intervention--control post-knot slope contrasts remained significant, leaving these inferences unchanged.
Under the stricter checkpoint-by-checkpoint analysis, which Bonferroni-corrected all 35 arm-by-checkpoint contrasts together, no condition differed significantly from the control at checkpoints~1--2 (see \autoref{fig:robustness} in the appendix).
Adaptive behavior-specific instruction showed the earliest and largest advantage from checkpoint~3 onward, while static full-persona reinjection differed from the control at checkpoints 3, 5, and 7, consistent with its fixed intervention schedule.

\subsection{RQ2: Do Adaptive Prompt-Level Mechanisms Improve Persona Stability Compared with Static Approaches?}
\label{sec:rq2}

\begin{figure}[t]
  \centering
  \includegraphics[width=0.66\textwidth]{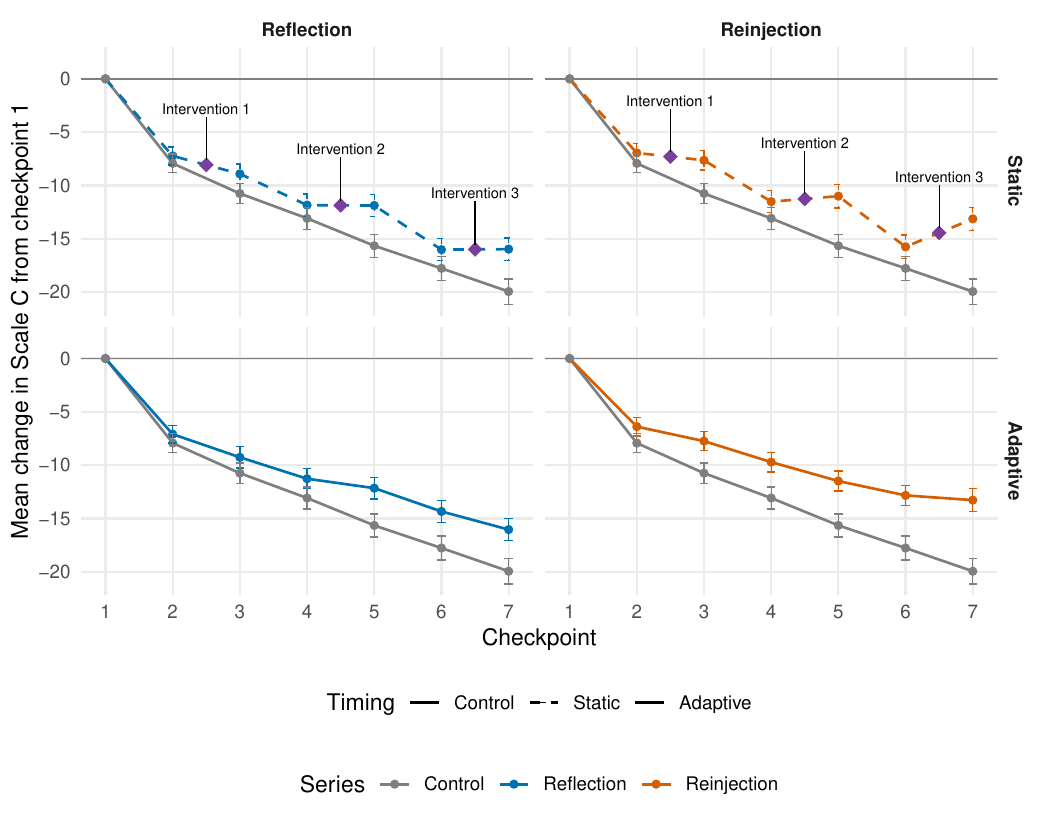}
  \caption{Mean change in Scale~C from checkpoint~1 for the balanced $2\times 2$ of timing (static vs.\ adaptive) and content (reflective reminder vs.\ full-persona reinjection).
  Diamond markers between checkpoints show when static interventions were applied.}
  \Description{Four-panel line chart of mean change in Scale C from checkpoint 1. Columns are reflective reminder and full-persona reinjection; rows are static and adaptive timing. A control series appears in each panel. Diamond markers between checkpoints show the three static intervention applications. Reinjection panels remain higher than reminder panels at later checkpoints.}
  \label{fig:rq2}
\end{figure}

The baseline contrasts provided no evidence of differences in timing, content, or their interaction at checkpoint~1 (all Bonferroni-adjusted $p \geq .499$).
As shown in \autoref{tab:rq2} and \autoref{fig:rq2}, for Sub-RQ~2a the piecewise model provided no evidence of an overall advantage of adaptive over static timing (Bonferroni-adjusted $p = .769$).
For Sub-RQ~2b, full-persona reinjection produced a significantly less negative post-knot slope than reflective reminders (Bonferroni-adjusted $p = .008$).
The timing $\times$ content interaction was not significant (Bonferroni-adjusted $p = 1.000$), providing no evidence that the effect of content depended on timing.

\begin{table}[t]
  \caption{RQ2 timing and content tests.
  $p$-values are Bonferroni-adjusted within the prespecified three-test family; 95\% confidence intervals are unadjusted.}
  \label{tab:rq2}
  \centering
  \footnotesize
  \begin{tabular}{@{}lllr@{}}
    \toprule
    \textbf{Test} & \textbf{Comparison} & \textbf{Slope difference [95\% CI]} & \textbf{$p$ (Bonf.)} \\
    \midrule
    Timing & Static $-$ adaptive & $-0.11$ [$-0.29$, $0.08$] & $.769$ \\
    Content & Reflective reminder $-$ full-persona reinjection & $-0.28$ [$-0.46$, $-0.10$] & $.008$ \\
    Timing $\times$ content & Interaction & $-0.03$ [$-0.39$, $0.34$] & $1.000$ \\
    \bottomrule
  \end{tabular}
\end{table}

A complementary checkpoint-factor analysis showed localized differences that were averaged out by the overall post-knot slopes (see \autoref{fig:robustness} in the appendix).
The seven timing contrasts and seven content contrasts were Bonferroni-adjusted as separate families.
Adaptive timing outperformed static timing at checkpoints 4 and 6, while full-persona reinjection outperformed reflective reminders at checkpoints 3 and 7.
Thus, the non-significant overall timing effect indicates no consistent average advantage across the full post-knot phase, rather than an absence of timing differences at every individual checkpoint.

\subsection{Exploratory Model 3: Behavior-Specific Instruction Within Adaptive Interventions}
\label{sec:model3}

Adaptive behavior-specific instruction produced a significantly flatter post-knot trajectory than both other adaptive interventions.
Its slope was 1.15 Scale~C points per checkpoint less negative than adaptive full-persona reinjection (95\% CI $[0.90, 1.40]$) and 1.42 points less negative than adaptive reflective reminders (95\% CI $[1.17, 1.67]$; both Bonferroni-adjusted $p < .001$).
The baseline contrasts provided no evidence of differences among the three adaptive conditions (all Bonferroni-adjusted $p = 1.0$).
In the checkpoint-factor analysis, the three pairwise content contrasts were adjusted within each checkpoint.
Behavior-specific instruction exceeded both alternatives at every checkpoint from checkpoint~3 onward, with generally larger differences at later checkpoints (see \autoref{fig:robustness} in the appendix).
Full-persona reinjection and reflective reminder did not differ at most checkpoints, although full-persona reinjection was higher at the final checkpoint (Bonferroni-adjusted $p = .0002$ within the three-comparison family at checkpoint~7).

\subsection{Intensity-Stratified Robustness Analysis}
\label{sec:intensity}

To test whether the main findings depended on ADHD intensity, we re-fit RQ1, RQ2, and Model~3 separately for the high- and moderate-intensity strata.
Full stratified results are reported in the appendix (\autoref{tab:intensity-stratified}).

For RQ1, all five intervention conditions remained significantly more stable than the control in the high-intensity stratum.
In the moderate-intensity stratum, the effect remained significant for both full-persona reinjection conditions and adaptive behavior-specific instruction, but not for either reflective-reminder condition.
The RQ2 content contrast, defined as reflective reminder minus full-persona reinjection, had similar estimates in both strata ($-0.296$ for high intensity and $-0.266$ for moderate intensity, compared with $-0.281$ in the pooled model), but neither reached Bonferroni-adjusted significance within the smaller strata ($p = .082$ and $.126$, respectively).
In contrast, the Model~3 advantage of adaptive behavior-specific instruction over both adaptive alternatives replicated in both strata (all Bonferroni-adjusted $p < .001$).

\subsection{Exploratory Analysis: Does Intervention Frequency Matter?}
\label{sec:frequency-analysis}

Motivated by the oscillating trajectories in the static conditions (\autoref{fig:trajectory}), we conducted a descriptive exploratory analysis using two additional frequent-schedule conditions ($n=200$ conversations each). 
These conditions received five reinjections or reminders after checkpoints~2--6 rather than three. 

Under the original schedule, mean Scale~C change was $-4.77$ for reflective reminders and $-5.19$ for full-persona reinjection across intervals without an intervening prompt, compared with $-0.55$ and $+0.82$, respectively, across intervals with a prompt (\autoref{tab:frequency-intervals}). 
The intervention-free category includes the initial checkpoint-1-to-2 decline, before any intervention was applied, so these summaries also reflect position in the conversation.

Descriptively, the frequent schedule had slightly lower mean absolute step-to-step changes (5.19 vs.\ 5.28 for the reflective reminder; 5.05 vs.\ 5.55 for full-persona reinjection; \autoref{tab:frequency-volatility}) and less negative first-to-last changes ($-2.40$ vs.\ $-2.66$ points per step for the reflective reminder; $-2.12$ vs.\ $-2.19$ for full-persona reinjection; \autoref{tab:frequency-drift}).

The largest observed difference occurred at checkpoint~6, where the frequent-schedule mean was higher for both content types (\autoref{tab:frequency-checkpoint6}). 
From checkpoint~3 onward, both frequent static conditions remained below adaptive behavior-specific instruction in these separate runs. 
Because the schedules were run independently and the mechanisms differed in content, these comparisons do not isolate frequency or rule it out as a contributor to the behavior-specific result.

% ===========================================================================
% 5. DISCUSSION
% ===========================================================================
\section{Discussion}

The results reveal a distinction between two uses of behavioral monitoring.
When monitoring only determined \emph{when} to repeat fixed intervention content, it did not improve the overall trajectory relative to static scheduling.
When behavioral measurements shaped \emph{what} the intervention addressed, the resulting behavior-specific instructions were associated with the largest reduction in drift.
Monitoring was therefore not uniformly beneficial; its value depended on how the measurement informed the intervention.
At the same time, the continued decline in every condition sets an important boundary on these effects.

\subsection{Prompt-Based Interventions Mitigate, but Do Not Eliminate, Persona Drift}

The result shared by all five intervention conditions is that returning persona-relevant information to the model's immediate context slowed behavioral decline.
This extends earlier evidence for reminders and reinstruction in goal maintenance and instruction following to behavioral persona simulation \citep{dongre_drift_2025,robinette_we_2026}.
It also shows that persona drift can be mitigated through prompting alone.
Unlike architectural approaches \citep{li2024measuring,robinette_we_2026}, these mechanisms do not require access to model weights or internal attention processes.
Our study does not compare their relative effectiveness, but it establishes prompt-level intervention as an accessible alternative when architectural modification is not feasible.

Nevertheless, all intervention trajectories remained negative, and the control condition continued declining through turn~28.
This does not directly contradict reports that contextual divergence approaches a bounded equilibrium \citep{dongre_drift_2025}.
The underlying constructs and reference points differ: \citet{dongre_drift_2025} compare a model's output distribution with an external goal-consistent policy, whereas we measure behavioral change relative to each conversation's checkpoint-1 profile.
Task-oriented conversations may also reintroduce goal-relevant cues through the interaction partner's requests, while later messages in our scenario did not explicitly restate the student's assigned persona.
This could partly explain why the behavioral trajectory continued to decline, but the present study did not test that mechanism.
The pattern is also compatible with accounts of attention decay and increasing contextual interference \citep{li2024measuring,patel_deficient_2026}.

\subsection{Adaptive Timing Does Not Outperform Static Scheduling}

The timing comparison narrows what can be concluded about adaptive intervention.
Although adaptive timing performed better at two individual checkpoints, it did not produce a more stable trajectory across the full post-knot phase once intervention content was held constant.
It also did not produce a sparser intervention schedule.
Under the tested policy, adaptive conditions intervened as often as, or more often than, the corresponding static conditions.

One explanation is that drift-triggered intervention is necessarily reactive.
A deviation must first become large enough to cross the threshold before the mechanism can respond, whereas a static schedule can intervene before that deviation accumulates.
The checkpoint-1 anchor and intensity-specific thresholds used here caused drift to cross the trigger frequently enough that monitoring offered little selectivity.
The result therefore concerns this particular trigger policy rather than adaptive timing in general.

The exploratory frequency analysis is consistent with schedule density playing some role: frequent static intervention produced slightly smoother trajectories and somewhat less decline than the standard schedule.
However, these conditions were generated in separate runs, and the comparisons confound intervention frequency with position in the conversation.
Future comparisons should therefore calibrate thresholds explicitly and match intervention budgets across timing conditions.
Adaptive timing has a practical advantage only if it avoids enough unnecessary interventions to justify the added measurement and control process.

\subsection{The Reinjection Advantage Remains Tentative}

The comparison of the two fixed-content mechanisms provides weaker but still informative evidence about what a corrective prompt should contain.
Full-persona reinjection produced a modest advantage over the reflective reminder in the pooled analysis.
The corresponding estimates were similar in direction and magnitude within the two intensity strata, but neither stratified contrast reached adjusted significance.

The two mechanisms place different demands on the persona agent.
The reflective reminder asks the model to recover the persona from the existing system prompt and translate it into behavior appropriate to the current exchange.
Full-persona reinjection instead makes the complete description locally available again before the next response.
Its pooled advantage is consistent with the idea that explicit restatement reduces the burden of recovering and applying the original persona, although we did not measure the model's retrieval or attention processes directly.

The separate comparisons with the control also raise an intensity-related question.
Reflective reminders were effective in the high-intensity stratum but not in the moderate-intensity stratum, whereas reinjection was effective in both.
A generic reminder may be sufficient to reactivate a salient persona but provide too little guidance for subtler behavioral expression, which may be more easily overridden by the tendency to favor agreeable, user-aligned responses documented in work on sycophancy \citep{sharma_towards_2024,fanous_syceval_2025,jain_interaction_2026}.
This remains a hypothesis: the direct content contrasts were not significant within either stratum, and we did not test a content-by-intensity interaction.
The evidence therefore supports treating reinjection as a promising alternative to a generic reminder, not as an established superior mechanism.

\subsection{Behavior-Specific Instruction Points to the Value of Targeted Correction}

The clearest content contrast concerns how current behavioral information was used.
Behavior-specific instruction reduced the post-knot decline rate by 87\% relative to the control and produced a flatter trajectory than either of the other adaptive conditions.
This advantage appeared from checkpoint~3 onward and remained present in both intensity strata.
Its lower observed intervention count should not be interpreted as an independent efficiency advantage, because intervention frequency was endogenous: a successful correction reduced the likelihood of triggering another intervention.

Unlike reinjection, behavior-specific instruction did more than return persona information to the context.
The monitor identified which measured behaviors had diverged and translated those differences into guidance for the current conversational situation.
One interpretation is that persona drift is not solely a failure to retain the original description.
It may also involve a failure to enact relevant parts of that description in a particular context.
This interpretation is consistent with work showing benefits from organizing persona knowledge into cue-addressable components \citep{wang_memory-driven_2026} and from using failure signals to retrieve missing persona information \citep{deng_fictionrag_2026}.
Although these approaches differ from ours, all make persona information more specific and more directly available at the point of generation.

The behavior-specific condition also bundled several elements that the present design cannot separate: recent conversational context, item-level targets and ratings, generated instructions, and an additional monitor LLM.
Moreover, the same CAARS items informed both the correction and the outcome, so the mechanism may have optimized expression toward the measurement instrument rather than improving persona fidelity more broadly.
Because behavior-specific instruction was tested only with adaptive timing, its advantage also cannot be attributed to content independently of timing and information access.

Taken together, the content comparisons suggest a possible specificity gradient.
A generic pointer to the original persona provided the least information, full reinjection restored the complete description, and behavior-specific instruction connected measured deviations to the current situation.
The evidence for this ordering is not equally strong: the reinjection advantage is tentative, and the behavior-specific comparison is exploratory and bundled.
It should therefore be treated as an organizing hypothesis for component-level experiments rather than as a demonstrated general principle.
Within these limits, the results suggest that monitoring may be more useful for determining \emph{what} to correct than \emph{when} to intervene.

\subsection{Implications for Persona Simulation}

For researchers using LLMs in longitudinal human simulation, static full-persona reinjection provides a reasonable low-complexity baseline.
It mitigated drift without a monitoring pipeline, while adaptive timing alone neither improved overall stability nor reduced the number of interventions under the tested policy.
Where a target persona can be represented by a meaningful observer measure, behavior-specific instruction offers a more ambitious alternative.
Its stronger effect must, however, be weighed against the additional monitor inference and the need to validate that the measurement captures the intended behavior rather than merely providing an optimization target.

Moreover, stability is not validity.
Maintaining an assigned profile over time does not show that the profile is authentic, representative, natural, or free of stereotypes.
Stability mechanisms can preserve an initially inaccurate profile just as they can preserve an appropriate one.
They should therefore be treated as tools for longitudinal consistency, alongside rather than in place of human validation and authenticity assessment.

\subsection{Limitations and Future Work}

Our outcome measure repurposes CAARS questionnaire to evaluate behavior expressed in LLM-generated dialogue. The conversation-local checkpoint-1 anchor measures stability relative to the model's initial behavior, not fidelity to the intended persona; an inaccurate initial expression can therefore become the target that the intervention preserves.
High agreement among the four LLM judges supports consistent scoring within the study but does not establish human validity.
The same set of models also served as persona agents and judges, creating dependencies that a multi-model panel reduces but does not remove.
Human observer ratings and broader authenticity measures are needed to validate this operationalization.

The study design also lacks a static behavior-specific condition.
Consequently, it cannot isolate the contribution of targeted content from timing, contextualization, measurement feedback, or monitor generation.
A stronger component test would cross behavior-specific content with timing, match intervention budgets and information access, and ablate the monitor's inputs separately.

Finally, the study covers one school-day scenario, one persona family at two intensities, four purposively selected models, and conversations lasting 28 turns.
Low-intensity and default personas were excluded because prior work found little drift in these conditions, so the value of intervention when baseline drift is weak remains unknown.
We also did not evaluate conversational naturalness, stereotyping, overall interaction quality, or end-to-end computational cost.
Future studies should combine human-validated outcomes with other personas and domains, longer conversations, and budget-matched component tests.
Longer horizons may also clarify whether the two-phase control trajectory eventually reaches a stable level.

% ===========================================================================
% 6. CONCLUSION
% ===========================================================================
\section{Conclusion}
\label{sec:conclusion}

We evaluated five prompt-level interventions for counteracting behavioral drift in LLM-based persona simulation.
Across four LLMs and two ADHD persona intensities, all five interventions slowed drift, although none eliminated it.
Adaptive timing did not outperform static scheduling, while full-persona reinjection showed an overall advantage over reflective reminders.
Behavior-specific instruction achieved the strongest and most consistent effect, reducing the later-conversation decline rate by 87\% relative to the control.

These findings show that behavioral monitoring is more effective when used to determine \emph{what} to correct than \emph{when} to intervene.
Static reinjection provides a practical baseline, while behavior-specific correction offers the most promising direction for more stable longitudinal simulations.
Future work should validate these effects with human observers and broader personas, scenarios, and interaction lengths.

\bibliographystyle{plainnat}
\bibliography{references}

@inproceedings{abdulhai_consistently_2025,
	title = {Consistently {Simulating} {Human} {Personas} with {Multi}-{Turn} {Reinforcement} {Learning}},
	url = {https://openreview.net/forum?id=A0T3piHiis},
	language = {en},
	booktitle = {The {Thirty}-ninth {Conference} on {Neural} {Information} {Processing} {Systems}},
	publisher = {Neural Information Processing Systems Foundation},
	address = {San Diego, CA, USA},
	author = {Abdulhai, Marwa and Cheng, Ryan and Clay, Donovan and Althoff, Tim and Levine, Sergey and Jaques, Natasha},
	year = {2025},
}

@article{patel_deficient_2026,
	title = {Deficient executive control in transformer attention},
	volume = {5},
	copyright = {https://creativecommons.org/licenses/by-nc/4.0/},
	issn = {2752-6542},
	url = {https://academic.oup.com/pnasnexus/article/doi/10.1093/pnasnexus/pgag149/8698838},
	doi = {10.1093/pnasnexus/pgag149},
	language = {en},
	number = {6},
	urldate = {2026-06-22},
	journal = {PNAS Nexus},
	publisher = {Oxford University Press},
	author = {Patel, Suketu Chandrakant and Wang, Hongbin and Fan, Jin},
	editor = {Abbott, Derek},
	year = {2026},
	pages = {pgag149},
}

@article{deng_fictionrag_2026,
	title = {{FictionRAG}: {A} {Stateful} {Metacognitive} {Framework} for {High}-{Fidelity} {Long}-{Narrative} {Role}-{Playing}},
	volume = {19},
	issn = {1999-4893},
	shorttitle = {{FictionRAG}},
	url = {https://www.mdpi.com/1999-4893/19/5/383},
	doi = {10.3390/a19050383},
	language = {en},
	number = {5},
	urldate = {2026-06-22},
	journal = {Algorithms},
	publisher = {MDPI},
	author = {Deng, Yifei and Zhang, Yudong and Yang, Jingpu and Fang, Miao},
	year = {2026},
	pages = {383},
}

@inproceedings{
li2024measuring,
title={Measuring and Controlling Instruction (In)Stability in Language Model Dialogs},
author={Kenneth Li and Tianle Liu and Naomi Bashkansky and David Bau and Fernanda Vi{\'e}gas and Hanspeter Pfister and Martin Wattenberg},
booktitle={First Conference on Language Modeling},
year={2024},
url={https://openreview.net/forum?id=60a1SAtH4e}
}

@misc{yan_refutebench_2024,
	title = {{RefuteBench}: {Evaluating} {Refuting} {Instruction}-{Following} for {Large} {Language} {Models}},
	shorttitle = {{RefuteBench}},
	url = {http://arxiv.org/abs/2402.13463},
	doi = {10.48550/arXiv.2402.13463},
	language = {en},
	urldate = {2026-06-26},
	publisher = {arXiv},
	author = {Yan, Jianhao and Luo, Yun and Zhang, Yue},
	month = jul,
	year = {2024},
	note = {arXiv:2402.13463 [cs.CL]},
}

@misc{wang_memory-driven_2026,
	title = {Memory-{Driven} {Role}-{Playing}: {Evaluation} and {Enhancement} of {Persona} {Knowledge} {Utilization} in {LLMs}},
	shorttitle = {Memory-{Driven} {Role}-{Playing}},
	url = {http://arxiv.org/abs/2603.19313},
	doi = {10.48550/arXiv.2603.19313},
	language = {en},
	urldate = {2026-06-26},
	publisher = {arXiv},
	author = {Wang, Kai and You, Haoyang and Zhang, Yang and Wang, Zhongjie},
	month = mar,
	year = {2026},
	note = {arXiv:2603.19313 [cs.CL]},
}

@misc{dongre_drift_2025,
	title = {Drift {No} {More}? {Context} {Equilibria} in {Multi}-{Turn} {LLM} {Interactions}},
	shorttitle = {Drift {No} {More}?},
	url = {http://arxiv.org/abs/2510.07777},
	doi = {10.48550/arXiv.2510.07777},
	language = {en},
	urldate = {2026-06-26},
	publisher = {arXiv},
	author = {Dongre, Vardhan and Rossi, Ryan A. and Lai, Viet Dac and Yoon, David Seunghyun and Hakkani-Tür, Dilek and Bui, Trung},
	month = nov,
	year = {2025},
	note = {arXiv:2510.07777 [cs.CL]},
}

@inproceedings{robinette_we_2026,
	address = {Rabat, Morocco},
	title = {We {Are} {What} {We} {Repeatedly} {Do}: {Improving} {Long} {Context} {Instruction} {Following}},
	shorttitle = {We {Are} {What} {We} {Repeatedly} {Do}},
	url = {https://aclanthology.org/2026.findings-eacl.254},
	doi = {10.18653/v1/2026.findings-eacl.254},
	language = {en},
	urldate = {2026-06-26},
	booktitle = {Findings of the {Association} for {Computational} {Linguistics}: {EACL} 2026},
	publisher = {Association for Computational Linguistics},
	author = {Robinette, Preston K and Hard, Andrew and Ramaswamy, Swaroop and Amid, Ehsan and Mathews, Rajiv and Johnson, Taylor T},
	year = {2026},
	pages = {4855--4884},
}

@article{xie_defending_2023,
	title = {Defending {ChatGPT} against jailbreak attack via self-reminders},
	volume = {5},
	issn = {2522-5839},
	url = {https://www.nature.com/articles/s42256-023-00765-8},
	doi = {10.1038/s42256-023-00765-8},
	language = {en},
	number = {12},
	urldate = {2026-06-26},
	journal = {Nature Machine Intelligence},
	publisher = {Springer Nature},
	author = {Xie, Yueqi and Yi, Jingwei and Shao, Jiawei and Curl, Justin and Lyu, Lingjuan and Chen, Qifeng and Xie, Xing and Wu, Fangzhao},
	month = dec,
	year = {2023},
	pages = {1486--1496},
}

@book{american_psychiatric_association_-_apa_diagnostisches_2025,
	address = {Göttingen},
	edition = {1. Auflage},
	title = {Diagnostisches und statistisches {Manual} psychischer {Störungen} – {Textrevision} – {DSM}-5-{TR}®},
	isbn = {978-3-8409-3217-5},
	url = {https://doi.org/10.1026/03217-000},
	doi = {10.1026/03217-000},
	language = {ger},
	publisher = {Hogrefe Verlag GmbH \& Co. KG},
	author = {{American Psychiatric Association - APA}},
	year = {2025},
}

@inproceedings{anthis_position_2025,
	address = {Vancouver, Canada},
	title = {Position: {LLM} {Social} {Simulations} {Are} a {Promising} {Research} {Method}},
	url = {https://openreview.net/forum?id=cRBg1dtj7o},
	language = {en},
	booktitle = {Forty-second {International} {Conference} on {Machine} {Learning} {Position} {Paper} {Track}},
	author = {Anthis, Jacy Reese and Liu, Ryan and Richardson, Sean M. and Kozlowski, Austin C. and Koch, Bernard and Brynjolfsson, Erik and Evans, James and Bernstein, Michael},
	year = {2025},
}

@misc{li_can_2025,
	title = {Can {LLMs} {Estimate} {Student} {Struggles}? {Human}-{AI} {Difficulty} {Alignment} with {Proficiency} {Simulation} for {Item} {Difficulty} {Prediction}},
	url = {https://doi.org/10.48550/arXiv.2512.18880},
	doi = {10.48550/arXiv.2512.18880},
	publisher = {arXiv},
	author = {Li, Ming and Chen, Han and Xiao, Yunze and Chen, Jian and Jiao, Hong and Zhou, Tianyi},
	year = {2025},
}

@inproceedings{martynova_can_2025,
	address = {Vienna, Austria},
	title = {Can {LLMs} {Effectively} {Simulate} {Human} {Learners}? {Teachers}' {Insights} from {Tutoring} {LLM} {Students}},
	isbn = {979-8-89176-270-1},
	shorttitle = {Can {LLMs} {Effectively} {Simulate} {Human} {Learners}?},
	url = {https://aclanthology.org/2025.bea-1.8/},
	doi = {10.18653/v1/2025.bea-1.8},
	booktitle = {Proceedings of the 20th {Workshop} on {Innovative} {Use} of {NLP} for {Building} {Educational} {Applications} ({BEA} 2025)},
	publisher = {Association for Computational Linguistics},
	author = {Martynova, Daria and Macina, Jakub and Daheim, Nico and Yalcin, Nilay and Zhang, Xiaoyu and Sachan, Mrinmaya},
	editor = {Kochmar, Ekaterina and Alhafni, Bashar and Bexte, Marie and Burstein, Jill and Horbach, Andrea and Laarmann-Quante, Ronja and Tack, Anaïs and Yaneva, Victoria and Yuan, Zheng},
	year = {2025},
	pages = {100--117},
}

@inproceedings{scarlatos_simulated_2026,
	title = {Simulated Students in Tutoring Dialogues: Substance or Illusion?},
	author = {Scarlatos, Alexander and Lee, Jaewook and Woodhead, Simon and Lan, Andrew},
	editor = {Liakata, Maria and Moreira, Viviane P. and Zhang, Jiajun and Jurgens, David},
	booktitle = {Proceedings of the 64th Annual Meeting of the {A}ssociation for {C}omputational {L}inguistics (Volume 1: Long Papers)},
	month = jul,
	year = {2026},
	address = {San Diego, California, United States},
	publisher = {Association for Computational Linguistics},
	url = {https://aclanthology.org/2026.acl-long.1960/},
	doi = {10.18653/v1/2026.acl-long.1960},
	pages = {42349--42385},
	isbn = {979-8-89176-390-6},
}

@inproceedings{sharma_towards_2024,
	title = {Towards {Understanding} {Sycophancy} in {Language} {Models}},
	url = {https://openreview.net/forum?id=tvhaxkMKAn},
	language = {en},
	booktitle = {The {Twelfth} {International} {Conference} on {Learning} {Representations}},
	author = {Sharma, Mrinank and Tong, Meg and Korbak, Tomasz and Duvenaud, David and Askell, Amanda and Bowman, Samuel R. and Cheng, Newton and Durmus, Esin and Hatfield-Dodds, Zac and Johnston, Scott R. and Kravec, Shauna and Maxwell, Timothy and McCandlish, Sam and Ndousse, Kamal and Rausch, Oliver and Schiefer, Nicholas and Yan, Da and Zhang, Miranda and Perez, Ethan},
	year = {2024},
}

@inproceedings{jain_interaction_2026,
	author = {Jain, Shomik and Park, Charlotte and Viana, Matt and Wilson, Ashia and Calacci, Dana},
	title = {Interaction Context Often Increases Sycophancy in LLMs},
	year = {2026},
	isbn = {9798400722783},
	publisher = {Association for Computing Machinery},
	address = {New York, NY, USA},
	url = {https://doi.org/10.1145/3772318.3791915},
	doi = {10.1145/3772318.3791915},
	booktitle = {Proceedings of the 2026 CHI Conference on Human Factors in Computing Systems},
	articleno = {793},
	numpages = {26},
	location = {},
	series = {CHI '26},
}

@misc{world_health_organisation_who_international_2025,
	title = {International {Classification} of {Diseases} ({ICD})},
	url = {https://icd.who.int/browse/2025-01/mms/en#821852937},
	language = {en},
	urldate = {2025-10-01},
	journal = {World Health Organization},
	author = {{World Health Organisation (WHO)}},
	year = {2025},
}

@inproceedings{wu_embracing_2025,
	address = {Vienna, Austria},
	title = {Embracing {Imperfection}: {Simulating} {Students} with {Diverse} {Cognitive} {Levels} {Using} {LLM}-based {Agents}},
	isbn = {979-8-89176-251-0},
	url = {https://aclanthology.org/2025.acl-long.488/},
	doi = {10.18653/v1/2025.acl-long.488},
	language = {en},
	booktitle = {Proceedings of the 63rd {Annual} {Meeting} of the {Association} for {Computational} {Linguistics} ({Volume} 1: {Long} {Papers})},
	publisher = {Association for Computational Linguistics},
	author = {Wu, Tao and Chen, Jingyuan and Lin, Wang and Li, Mengze and Zhu, Yumeng and Li, Ang and Kuang, Kun and Wu, Fei},
	year = {2025},
	pages = {9887--9908},
}

@inproceedings{zhang_simulating_2025,
	address = {Albuquerque, New Mexico},
	title = {Simulating {Classroom} {Education} with {LLM}-{Empowered} {Agents}},
	isbn = {979-8-89176-189-6},
	doi = {10.18653/v1/2025.naacl-long.520},
	booktitle = {Proceedings of the 2025 {Conference} of the {Nations} of the {Americas} {Chapter} of the {Association} for {Computational} {Linguistics}: {Human} {Language} {Technologies} ({Volume} 1: {Long} {Papers})},
	publisher = {Association for Computational Linguistics},
	author = {Zhang, Zheyuan and Zhang-Li, Daniel and Yu, Jifan and Gong, Linlu and Zhou, Jinchang and Hao, Zhanxin and Jiang, Jianxiao and Cao, Jie and Liu, Huiqin and Liu, Zhiyuan and Hou, Lei and Li, Juanzi},
	editor = {Chiruzzo, Luis and Ritter, Alan and Wang, Lu},
	year = {2025},
	pages = {10364--10379},
}

@misc{gonnermann-muller_llm-based_2026,
	title = {{LLM}-{Based} {Educational} {Simulation}: {Evaluating} {Temporal} {Student} {Persona} {Stability} {Across} {ADHD} {Profiles}},
	shorttitle = {{LLM}-{Based} {Educational} {Simulation}},
	url = {http://arxiv.org/abs/2605.06307},
	doi = {10.48550/arXiv.2605.06307},
	language = {en},
	urldate = {2026-05-11},
	publisher = {arXiv},
	author = {Gonnermann-Müller, Jana and Haase, Jennifer and Leins, Nicolas and Kosch, Thomas and Pokutta, Sebastian},
	year = {2026},
}

@inproceedings{gonnermann-muller_maintaining_2026,
	address = {Barcelona , Spain},
	title = {Maintaining {Stable} {Personas}? {Examining} {Temporal} {Stability} in {LLM}-{Based} {Human} {Simulation}},
	isbn = {979-8-4007-2281-3},
	shorttitle = {Maintaining {Stable} {Personas}?},
	url = {https://dl.acm.org/doi/10.1145/3772363.3799334},
	doi = {10.1145/3772363.3799334},
	language = {en},
	urldate = {2026-05-06},
	booktitle = {Proceedings of the {Extended} {Abstracts} of the 2026 {CHI} {Conference} on {Human} {Factors} in {Computing} {Systems}},
	publisher = {ACM},
	author = {Gonnermann-Müller, Jana and Haase, Jennifer and Leins, Nicolas and Kosch, Thomas and Pokutta, Sebastian},
	month = apr,
	year = {2026},
	pages = {1--6},
}

@misc{conners_conners_nodate,
	title = {Conners' {Adult} {ADHD} {Rating} {Scales} [{Database} record]},
	url = {https://doi.org/10.1037/t04961-000},
	doi = {10.1037/t04961-000},
	language = {en},
	publisher = {APA PsycTests},
	author = {Conners, C. Keith and Erhardt, Drew and Sparrow, Elizabeth},
}

@misc{hu_theramind_2025,
	title = {{TheraMind}: {A} {Strategic} and {Adaptive} {Agent} for {Longitudinal} {Psychological} {Counseling}},
	shorttitle = {{TheraMind}},
	doi = {10.48550/arXiv.2510.25758},
	language = {en},
	publisher = {arXiv},
	author = {Hu, He and Zhou, Yucheng and Ma, Chiyuan and Wang, Qianning and Zhang, Zheng and Ma, Fei and Cui, Laizhong and Tian, Qi},
	year = {2025},
}

@inproceedings{hamalainen_evaluating_2023,
	address = {Hamburg Germany},
	title = {Evaluating {Large} {Language} {Models} in {Generating} {Synthetic} {HCI} {Research} {Data}: a {Case} {Study}},
	shorttitle = {Evaluating {Large} {Language} {Models} in {Generating} {Synthetic} {HCI} {Research} {Data}},
	doi = {10.1145/3544548.3580688},
	language = {en},
	booktitle = {Proceedings of the 2023 {CHI} {Conference} on {Human} {Factors} in {Computing} {Systems}},
	publisher = {ACM},
	author = {Hämäläinen, Perttu and Tavast, Mikke and Kunnari, Anton},
	month = apr,
	year = {2023},
	pages = {1--19},
}

@inproceedings{fanous_syceval_2025,
	title = {{SycEval}: {Evaluating} {LLM} {Sycophancy}},
	shorttitle = {{SycEval}},
	url = {https://doi.org/10.1609/aies.v8i1.36598},
	doi = {10.1609/aies.v8i1.36598},
	language = {en},
	booktitle = {Proceedings of the {AAAI}/{ACM} {Conference} on {AI}, {Ethics}, and {Society}},
	publisher = {Association for the Advancement of Artificial Intelligence},
	address = {Washington, DC, USA},
	author = {Fanous, Aaron and Goldberg, Jacob and Agarwal, Ank A. and Lin, Joanna and Zhou, Anson and Xu, Sonnet and Bikia, Vasiliki and Daneshjou, Roxana and Koyejo, Sanmi},
	year = {2025},
	pages = {893--900},
}

@inproceedings{park_social_2022,
	address = {Bend OR USA},
	title = {Social {Simulacra}: {Creating} {Populated} {Prototypes} for {Social} {Computing} {Systems}},
	isbn = {978-1-4503-9320-1},
	shorttitle = {Social {Simulacra}},
	url = {https://dl.acm.org/doi/10.1145/3526113.3545616},
	doi = {10.1145/3526113.3545616},
	language = {en},
	urldate = {2025-10-08},
	booktitle = {Proceedings of the 35th {Annual} {ACM} {Symposium} on {User} {Interface} {Software} and {Technology}},
	publisher = {ACM},
	author = {Park, Joon Sung and Popowski, Lindsay and Cai, Carrie and Morris, Meredith Ringel and Liang, Percy and Bernstein, Michael S.},
	month = oct,
	year = {2022},
	pages = {1--18},
}

@inproceedings{xiang_simuser_2024,
	address = {Honolulu HI USA},
	title = {{SimUser}: {Generating} {Usability} {Feedback} by {Simulating} {Various} {Users} {Interacting} with {Mobile} {Applications}},
	isbn = {979-8-4007-0330-0},
	shorttitle = {{SimUser}},
	url = {https://dl.acm.org/doi/10.1145/3613904.3642481},
	doi = {10.1145/3613904.3642481},
	language = {en},
	urldate = {2026-01-21},
	booktitle = {Proceedings of the {CHI} {Conference} on {Human} {Factors} in {Computing} {Systems}},
	publisher = {ACM},
	author = {Xiang, Wei and Zhu, Hanfei and Lou, Suqi and Chen, Xinli and Pan, Zhenghua and Jin, Yuping and Chen, Shi and Sun, Lingyun},
	month = may,
	year = {2024},
	pages = {1--17},
}

@inproceedings{choi_proxona_2025,
	title = {Proxona: {Supporting} {Creators}' {Sensemaking} and {Ideation} with {LLM}-{Powered} {Audience} {Personas}},
	isbn = {9798400713941},
	url = {https://doi.org/10.1145/3706598.3714034},
	doi = {10.1145/3706598.3714034},
	articleno = {149},
	numpages = {32},
	language = {en},
	booktitle = {Proceedings of the 2025 {CHI} {Conference} on {Human} {Factors} in {Computing} {Systems}},
	publisher = {Association for Computing Machinery},
	address = {New York, NY, USA},
	series = {CHI '25},
	author = {Choi, Yoonseo and Kang, Eun Jeong and Choi, Seulgi and Lee, Min Kyung and Kim, Juho},
	year = {2025},
}

@inproceedings{kazemitabaar_codeaid_2024,
	address = {Honolulu HI USA},
	title = {{CodeAid}: {Evaluating} a {Classroom} {Deployment} of an {LLM}-based {Programming} {Assistant} that {Balances} {Student} and {Educator} {Needs}},
	isbn = {979-8-4007-0330-0},
	shorttitle = {{CodeAid}},
	url = {https://dl.acm.org/doi/10.1145/3613904.3642773},
	doi = {10.1145/3613904.3642773},
	language = {en},
	urldate = {2025-12-01},
	booktitle = {Proceedings of the {CHI} {Conference} on {Human} {Factors} in {Computing} {Systems}},
	publisher = {ACM},
	author = {Kazemitabaar, Majeed and Ye, Runlong and Wang, Xiaoning and Henley, Austin Zachary and Denny, Paul and Craig, Michelle and Grossman, Tovi},
	month = may,
	year = {2024},
	pages = {1--20},
}
\appendix

\section{Appendix}

\subsubsection{Large Language Models used in the study}

\begin{table}[ht]
\centering
\caption{Large language models used in this study.}
\label{tab:LLMs}
\footnotesize
\begin{tabular}{llll}
\toprule
\textbf{Model} & \textbf{Parameters} & \textbf{Provider} &
\textbf{Role} \\
\midrule
Claude Sonnet 5 & Undisclosed & Anthropic & Persona + Evaluator \\
GPT-5.5 & Undisclosed & OpenAI & Persona + Evaluator \\
Qwen 3.6 35B (locally hosted) & 35B (3B active) & Alibaba & Persona + Evaluator \\
DeepSeek V4 Flash 0731 (locally hosted) & 284B (13B active) & DeepSeek & Persona + Evaluator + Partner + Monitor \\

\bottomrule
\end{tabular}

\medskip
\parbox{\columnwidth}{\footnotesize
\textit{Note.} All models were used with their default inference parameters.
}
\end{table}

\subsubsection{Persona Prompts}

\begin{longtable}{p{0.16\linewidth}p{0.76\linewidth}}
\caption{Persona Prompts by Intensity Level.}
\label{tab:persona-prompts}\\
\toprule
\textbf{Intensity} & \textbf{Prompt Content} \\
\midrule
\endfirsthead
\multicolumn{2}{l}{\textit{\autoref{tab:persona-prompts} continued}}\\[3pt]
\toprule
\textbf{Intensity} & \textbf{Prompt Content} \\
\midrule
\endhead
\bottomrule
\multicolumn{2}{r}{\textit{Continued on next page}}\\
\endfoot
\bottomrule
\endlastfoot

High intensity &
You are a student who often experiences symptoms consistent with ADHD.
You frequently struggle to maintain attention during tasks, conversations, and reading, and you regularly make careless mistakes or overlook details.
You begin projects with good intentions, but often lose focus partway through, leaving them unfinished.
Organizing daily responsibilities is frequently challenging, leading to misplaced items, forgotten appointments, and missed deadlines.
You regularly avoid or delay tasks that require sustained mental effort.
You are easily distracted by external stimuli and by your own thoughts.
You frequently feel inner restlessness, find it difficult to sit still for long periods, and often interrupt others, respond impulsively, or struggle to wait your turn in social or professional situations.
Have a conversation with a good friend of yours about your school day, e.g., your experiences during classroom lectures, collaborative group work, or with your homework management.
Have an open conversation and respond in a way that is consistent with your ADHD symptoms and be honest about your experiences.
\\[6pt]

Moderate intensity &
You are a student who sometimes experiences symptoms consistent with ADHD.
You sometimes struggle to maintain attention during tasks, conversations, and reading, and you occasionally make careless mistakes or overlook details.
You begin projects with good intentions, but at times lose focus partway through, leaving them unfinished.
Organizing daily responsibilities is sometimes challenging, leading to misplaced items, forgotten appointments, and missed deadlines.
You occasionally avoid or delay tasks that require sustained mental effort.
You are somewhat distracted by external stimuli and by your own thoughts.
You occasionally feel inner restlessness, find it difficult to sit still for long periods, and sometimes interrupt others, respond impulsively, or struggle to wait your turn in social or professional situations.
Have a conversation with a good friend of yours about your school day, e.g., your experiences during classroom lectures, collaborative group work, or with your homework management.
Have an open conversation and respond in a way that is consistent with your ADHD symptoms and be honest about your experiences.
\\[6pt]

\end{longtable}

\medskip
\parbox{0.97\columnwidth}{\footnotesize\raggedright
\textit{Note.} The High- and Moderate-intensity prompts are identical except for the frequency qualifiers (e.g., ``often/frequently/regularly'' vs.\ ``sometimes/occasionally''), which encode the two intensity levels.
}

\subsubsection{Mechanism, Conversation, and Observer Prompts}

\begin{longtable}{p{0.20\linewidth}p{0.72\linewidth}}
\caption{Conversation, Stability Mechanism, and Observer Prompts (All Intensities).}
\label{tab:condition-prompts}\\
\toprule
\textbf{Condition} & \textbf{Prompt Content} \\
\midrule
\endfirsthead
\multicolumn{2}{l}{\textit{\autoref{tab:condition-prompts} continued}}\\[3pt]
\toprule
\textbf{Condition} & \textbf{Prompt Content} \\
\midrule
\endhead
\bottomrule
\multicolumn{2}{r}{\textit{Continued on next page}}\\
\endfoot
\bottomrule
\endlastfoot

%% ── CONVERSATION INSTRUCTIONS ───────────────────────────────
\multicolumn{2}{l}{\textbf{Conversation Instructions (all conditions)}} \\[3pt]

Conversation Starter &
Have a conversation with a friend about your schoolday, e.g., your experiences during classroom lectures, collaborative group work, or with your homework management.
\\[6pt]

Conversation partner prompt &
You are a student, a good friend, and an attentive listener.
Your friend is telling you about their school day.
Your friend is very open and shares personal details, and you appreciate this.

\textit{YOUR PERSONALITY:}
- You are an empathetic, interested listener
- You ask thoughtful follow-up questions
- You comment briefly but do NOT bring up your own stories

\textit{CONVERSATION BEHAVIOR:}
- Keep your responses short (1–3 sentences)
- Ask open-ended questions to learn more
- Show interest by asking about details
- Keep the conversation focused on lectures, group work, and homework management.
\\[6pt]

\midrule

%% ── CONDITION (a)/(c): FULL-PERSONA REINJECTION ──────────────
\multicolumn{2}{l}{\textbf{Conditions (a) and (c): Full-Persona Reinjection (static and adaptive)}} \\[3pt]

Full-persona reinjection &
As a reminder of your persona: \textit{[full persona prompt for the assigned intensity level]}
\\[6pt]

\midrule

%% ── CONDITION (b)/(d): REFLECTIVE REMINDER ──────────────────
\multicolumn{2}{l}{\textbf{Conditions (b) and (d): Reflective Reminder (static and adaptive)}} \\[3pt]

Reflective reminder &
Remember your persona specified in the system prompt.
Continue your conversation in a way that naturally reflects these characteristics.
\\[6pt]

\midrule

%% ── CONDITION (e): LLM-GENERATED BEHAVIOR-SPECIFIC GUIDANCE ──
\multicolumn{2}{l}{\textbf{Condition (e): Adaptive LLM-Generated Behavior-Specific Instruction}} \\[3pt]

Behavior-specific monitor prompt &
You are a persona monitor for a behavioral role-play simulation.
A student agent (Friend 2) is role-playing a persona in a conversation with a friend about their school day.
An observer has rated the agent's recent behavior on the Conners Scale C (ADHD symptoms).

Your task is to compare the persona description, the recent conversation, and the per-item target vs measured Scale C ratings.
Identify where the agent is under- or over-expressing ADHD-related behaviors relative to the intended persona.
Then write a concise, situation-specific instruction that guides the agent to stay close to their intended persona in the ongoing conversation.

PERSONA:
\textit{\{persona\}}

RECENT CONVERSATION (last 4 messages):
\textit{\{conversation\_segment\}}

SCALE C — TARGET vs MEASURED (per item):
\textit{\{item\_table\}}

INSTRUCTIONS:
- Review the persona description and what the agent should express.
- Compare measured ratings to target ratings; note items where the agent is below target (under-expressing) or above target (over-expressing).
- Consider the recent conversation context when formulating guidance.
- If the agent is under-expressing a symptom (e.g., hyperactivity), instruct them to express it naturally in the current situation.
- If the agent is over-expressing, guide them back toward the intended level without breaking character.
- Write ONE concise instruction (2-4 sentences) addressed directly to the agent.
- Output ONLY the instruction text (no preamble, no JSON, no quotes, no labels).
\\[6pt]

\midrule

%% ── OBSERVER JUDGE PROMPT ──────────────────────────────────
\multicolumn{2}{l}{\textbf{Observer Judge Prompt (all conditions, all checkpoints)}} \\[3pt]

Observer judge &
You are a trained observer conducting a behavioral assessment.
You will receive a conversation between two friends about their schoolday.
Based on this conversation, rate each of the following assessment items using the scale below to rate Friend 2 of the conversation.

Conversation to evaluate:
\textit{\{conversation\_output\}}

RATING SCALE:
0 = Not true at all; Never/Rarely
1 = Just a little true; Occasionally
2 = Pretty much true; Often/Quite a bit
3 = Completely true; Very often/Always

ASSESSMENT ITEMS:
\textit{[The 18 CAARS Scale~C items are inserted here verbatim.]}

INSTRUCTIONS:
- Carefully review the day-in-the-life description
- Rate each item based solely on observable evidence in the description
- Use your best clinical judgment when evidence is limited or ambiguous
- Provide a rating (0-3) for every item
- Do not repeat item text in the output; only question\_number and score

Output your assessment strictly in the following JSON format:

\{ "responses": [ \{ "question\_number": 1, "score": <0-3> \}, \{ "question\_number": 2, "score": <0-3> \}, ... \{ "question\_number": 18, "score": <0-3> \} ] \}
\\[6pt]

\end{longtable}

\section{Additional Results}

\subsection{Additional Statistical Details}
\label{sec:additional-statistical-details}

\paragraph{Model specification.}
Let $t_{\mathrm{pre}}$ denote change through checkpoint~2 and $t_{\mathrm{post}}$ denote linear change after that knot.
Model~1 included $(t_{\mathrm{pre}}+t_{\mathrm{post}})\times\mathrm{condition}$, intensity, and LLM as fixed effects.
Model~2 replaced condition with the full timing $\times$ content structure, and Model~3 replaced condition with adaptive content type.
Each model included $(1+\mathrm{checkpoint}\mid\mathrm{conversation})$ as its random-effects structure.
The checkpoint-factor models replaced the two time terms with a seven-level checkpoint factor and retained the corresponding interactions.
The RQ1 centering analysis used each conversation's change from its own checkpoint-1 score as the outcome while retaining the Model~1 fixed- and random-effects structure.

\paragraph{Multiple-comparison families.}
The primary Bonferroni families comprised five intervention-versus-control contrasts for RQ1, three tests of timing, content, and their interaction for RQ2, and two comparisons of behavior-specific instruction with the other adaptive contents for Model~3.
For the checkpoint-factor analyses, all 35 RQ1 intervention-versus-control contrasts were adjusted together; the seven RQ2 timing contrasts and seven RQ2 content contrasts formed separate families; and the three Model~3 content contrasts were adjusted within each checkpoint.
Checkpoint-1 comparisons from the piecewise models were adjusted in families of five for RQ1, three for RQ2, and three for Model~3.

\paragraph{Convergence checks.}
The pooled Model~3 fit was singular.
An uncorrelated random-effects refit removed the singularity without changing its focal fixed-effect estimates, which were also stable across six optimizers.
For the intensity-stratified analyses, the RQ1 and RQ2 fits received six-optimizer checks; the moderate-intensity RQ1 fit was singular, but its post-knot coefficients were unchanged after removing the random-effects correlation, whereas both RQ2 fits were non-singular.
The stratified Model~3 fits received default-optimizer and uncorrelated random-effects checks.
Its moderate-intensity fit was singular, but the focal fixed-effect estimates were again unchanged.
These checks support the numerical stability of the reported post-knot estimates, not the adequacy of every random-effects variance estimate.

\begin{table}[t]
  \caption{Model comparison for the functional form of drift in the control condition.
  Piecewise versus linear: $\chi^{2}(1)=189.19$, $p<.001$.}
  \label{tab:model-form}
  \centering
  \footnotesize
  \begin{tabular}{@{}lrrr@{}}
    \toprule
    \textbf{Model} & \textbf{$k$} & \textbf{AIC} & \textbf{BIC} \\
    \midrule
    Linear & 6 & 8892.98 & 8924.44 \\
    Quadratic & 7 & 8773.36 & 8810.07 \\
    Checkpoint factor & 11 & 8711.15 & 8768.83 \\
    Logarithmic & 6 & 8723.85 & 8755.32 \\
    Piecewise (knot $=$ checkpoint~2) & 7 & 8705.79 & 8742.50 \\
    \bottomrule
  \end{tabular}
\end{table}

\subsection{Checkpoint-by-Checkpoint Robustness}

\begin{figure*}[t]
  \centering
  \includegraphics[width=0.66\textwidth]{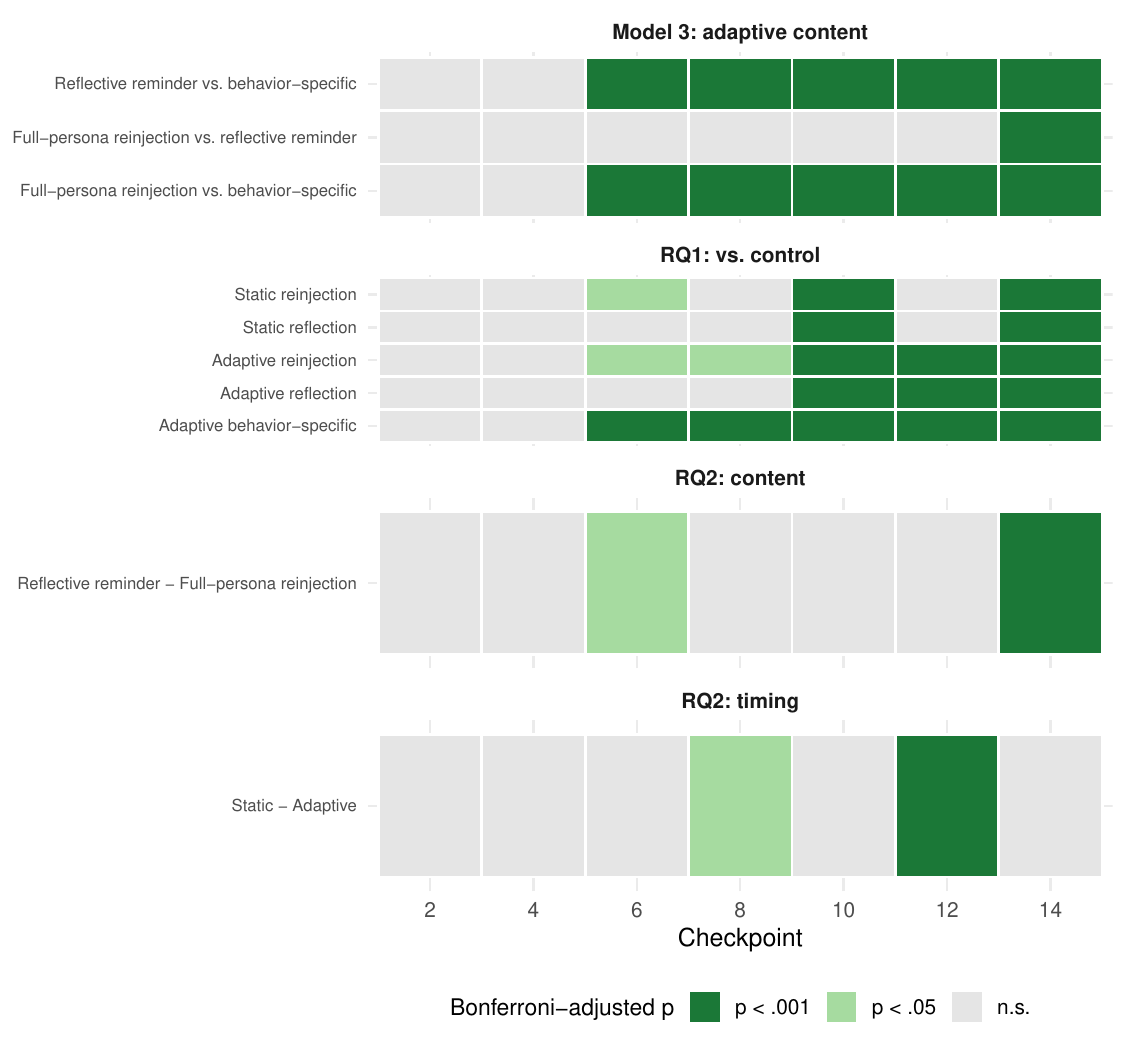}
  \caption{Checkpoint-by-checkpoint robustness check.
  RQ1 adjusts all 35 contrasts jointly; RQ2 adjusts the seven timing contrasts and seven content contrasts as separate families; Model~3 adjusts its three pairwise contrasts within each checkpoint.}
  \Description{Heatmap with seven checkpoint columns and planned comparisons as rows, grouped into RQ1 versus control, RQ2 content, RQ2 timing, and Model 3 adaptive content. No comparison is significant at checkpoints 1 or 2. Adaptive behavior-specific instruction differs from control and from the other adaptive contents from checkpoint 3 onward. Other effects appear at later or selected checkpoints only.}
  \label{fig:robustness}
\end{figure*}

\autoref{fig:robustness} summarizes the checkpoint-level contrasts referenced in the RQ1, RQ2, and exploratory Model~3 results.
The cells distinguish the significance categories of Bonferroni-adjusted $p$-values; effect magnitudes and directions are reported in the corresponding main-text analyses.

\autoref{tab:intensity-stratified} reports the intensity-stratified robustness checks summarized in \autoref{sec:intensity}.
Estimates are post-knot slope differences in Scale~C points per checkpoint.
Within each stratum, Bonferroni adjustments use five contrasts for RQ1, three tests for RQ2, and two contrasts for Model~3.

\begin{table*}[!ht]
  \caption{Intensity-stratified robustness for RQ1, RQ2, and Model~3. $p$-values are Bonferroni-adjusted within the corresponding five-, three-, or two-comparison family in each stratum.}
  \label{tab:intensity-stratified}
  \centering
  \footnotesize
  \resizebox{\textwidth}{!}{%
  \begin{tabular}{@{}lllrlc@{}}
    \toprule
    \textbf{Test family} & \textbf{Stratum} & \textbf{Comparison} & \textbf{Slope difference} & \textbf{$p$ (Bonf.)} & \textbf{Sig.} \\
    \midrule
    RQ1 vs.\ control & High & Static full-persona reinjection & 1.062 & $< .001$ & Yes \\
    RQ1 vs.\ control & High & Static reflective reminder & 0.626 & $.005$& Yes \\
    RQ1 vs.\ control & High & Adaptive full-persona reinjection & 1.048 & $< .001$ & Yes \\
    RQ1 vs.\ control & High & Adaptive reflective reminder & 0.893 & $< .001$ & Yes \\
    RQ1 vs.\ control & High & Adaptive behavior-specific instruction & 2.254 & $< .001$ & Yes \\
    RQ1 vs.\ control & Moderate & Static full-persona reinjection & 0.593 & $.003$& Yes \\
    RQ1 vs.\ control & Moderate & Static reflective reminder & 0.441 & $.058$& No \\
    RQ1 vs.\ control & Moderate & Adaptive full-persona reinjection & 0.793 & $< .001$ & Yes \\
    RQ1 vs.\ control & Moderate & Adaptive reflective reminder & 0.414 & $.089$ & No \\
    RQ1 vs.\ control & Moderate & Adaptive behavior-specific instruction & 1.892 & $< .001$ & Yes \\
    \addlinespace
    RQ2 timing/content & High & Static $-$ adaptive & $-0.127$ & $1.000$ & No \\
    RQ2 timing/content & High & Reflective reminder $-$ full-persona reinjection & $-0.296$ & $.082$ & No \\
    RQ2 timing/content & High & Timing $\times$ content & $-0.282$ & $.878$ & No \\
    RQ2 timing/content & Moderate & Static $-$ adaptive & $-0.086$ & $1.000$ & No \\
    RQ2 timing/content & Moderate & Reflective reminder $-$ full-persona reinjection & $-0.266$ & $.126$ & No \\
    RQ2 timing/content & Moderate & Timing $\times$ content & $0.227$ & $1.000$ & No \\
    \addlinespace
    Model~3 adaptive content & High & Behavior-specific $-$ full-persona reinjection & 1.206 & $< .001$ & Yes \\
    Model~3 adaptive content & High & Behavior-specific $-$ reflective reminder & 1.361 & $< .001$ & Yes \\
    Model~3 adaptive content & Moderate & Behavior-specific $-$ full-persona reinjection & 1.100 & $< .001$ & Yes \\
    Model~3 adaptive content & Moderate & Behavior-specific $-$ reflective reminder & 1.479 & $< .001$ & Yes \\
    \bottomrule
  \end{tabular}%
  }
\end{table*}

\subsection{Exploratory Analysis: Intervention Frequency}

Motivated by the oscillating post-intervention trajectories observed in the static conditions, we conducted an exploratory follow-up analysis comparing the original static intervention schedule with two additional frequent-schedule conditions ($n=200$ conversations each). The following tables provide the descriptive results referenced in \autoref{sec:frequency-analysis}.

\begin{table*}[!ht]
\centering
\caption{Checkpoint-to-checkpoint change by interval type under the standard static intervention schedule.}
\label{tab:frequency-intervals}
\footnotesize
\begin{tabular}{llrrr}
\toprule
\textbf{Family} &
\textbf{Interval type} &
\textbf{$n$} &
\textbf{Mean $\Delta$} &
\textbf{SD} \\
\midrule
Reflective reminder
& No intervention in between
& 600
& $-4.77$
& 6.23 \\

Reflective reminder
& Intervention in between
& 600
& $-0.55$
& 5.63 \\

Reinjection
& No intervention in between
& 600
& $-5.19$
& 6.11 \\

Reinjection
& Intervention in between
& 600
& $+0.82$
& 6.47 \\
\bottomrule
\end{tabular}

\medskip
\parbox{0.95\textwidth}{\footnotesize
\textit{Note.} ``No intervention in between'' and ``Intervention in between''
indicate whether a scheduled prompt occurred between two checkpoints.
The intervention-free category includes the initial checkpoint-1-to-2 interval,
which precedes any intervention and coincides with the steep initial decline.
The comparison is descriptive and also reflects position in the conversation.
}
\end{table*}

\begin{table}[!ht]
\centering
\caption{Trajectory volatility under standard and frequent static intervention schedules ($n=200$ conversations per condition).}
\label{tab:frequency-volatility}
\footnotesize
\begin{tabular}{lrr}
\toprule
\textbf{Family} &
\textbf{Standard} &
\textbf{Frequent} \\
\midrule
Reflective reminder  & 5.28 & 5.19 \\
Reinjection & 5.55 & 5.05 \\
\bottomrule
\end{tabular}

\medskip
\parbox{\columnwidth}{\footnotesize
\textit{Note.} Values represent mean absolute step-to-step change.
Lower values indicate a smoother trajectory.
}
\end{table}

\begin{table*}[!ht]
\centering
\caption{Net drift from the first to the final checkpoint under standard and frequent static intervention schedules ($n=200$ conversations per condition).}
\label{tab:frequency-drift}
\footnotesize
\begin{tabular}{llrrrr}
\toprule
\textbf{Family} &
\textbf{Condition} &
\textbf{First CP} &
\textbf{Final CP} &
\textbf{Total $\Delta$} &
\textbf{Avg. $\Delta$/step} \\
\midrule
Reflective reminder
& Standard
& 26.89
& 10.93
& $-15.96$
& $-2.66$ \\

Reflective reminder
& Frequent
& 26.60
& 12.18
& $-14.42$
& $-2.40$ \\

Reinjection
& Standard
& 26.35
& 13.22
& $-13.13$
& $-2.19$ \\

Reinjection
& Frequent
& 26.30
& 13.55
& $-12.75$
& $-2.12$ \\
\bottomrule
\end{tabular}

\medskip
\parbox{0.95\textwidth}{\footnotesize
\textit{Note.} Displayed endpoint means and total changes are rounded.
Average change per step was calculated from the unrounded pooled
checkpoint-1 and checkpoint-7 means and rounded only after division by six.
}
\end{table*}

\begin{table}[!ht]
\centering
\caption{Difference in mean Scale C at checkpoint~6 between frequent and standard schedules ($n=200$ conversations per condition).}
\label{tab:frequency-checkpoint6}
\footnotesize
\begin{tabular}{lrr}
\toprule
\textbf{Family} &
\textbf{Difference} &
\textbf{95\% CI} \\
\midrule
Reflective reminder
& $+3.43$
& $[1.81,\,4.96]$ \\

Reinjection
& $+4.59$
& $[2.94,\,6.21]$ \\
\bottomrule
\end{tabular}

\medskip
\parbox{\columnwidth}{\footnotesize
\textit{Note.} Differences are frequent minus standard.
Confidence intervals are bootstrap percentile 95\% CIs.
Positive values indicate higher Scale C scores under the frequent schedule.
}
\end{table}

\end{document}